# Investigation of activation cross sections for deuteron induced reactions on strontium up to 50 MeV

**F. Tárkányi[a], A. Hermanne[b], F. Ditrói[a*], S. Takács[a], Z. Szücs[a], K. Brezovcsik[a]**

a Institute for Nuclear Research, Hungarian Academy of Sciences (ATOMKI), 4026 Debrecen, Hungary

b Cyclotron Laboratory, Vrije Universiteit Brussel (VUB), Laarbeeklaan 103, 1090 Brussels, Belgium

## *Abstract*

Excitation functions were measured for the $^{nat}Sr(d,x)^{88,87m,87g,86g,85g}Y$, $^{87m,85g,83g,82}Sr$, $^{86g,84g,83,82m,81g}Rb$ reactions by the stacked foil activation technique and high-resolution gamma-spectrometry up to 50 MeV. We present the first experimental activation cross section data for all investigated reactions. Our experimental data are compared with the TALYS code results as available in the TENDL-2015 on-line library. Use of deuteron induced reactions on Sr for production of medical isotopes is discussed.

* Corresponding author: ditroi@atomki.hu

## *Introduction*

We are performing a systematic study of activation cross- sections of light charged particle induced reactions. The aim is to get reliable experimental data for optimizing production of specific radionuclides for different applications and to consolidate the database allowing improvement of theoretical models. In the present work we report experimental cross section data for deuteron induced reactions on strontium. The investigated activation radio-products are of interest in different applications:

**$^{90}$Y** is widely used in brachytherapy and radioimmunotherapy but does not emit photons for imaging (Wiseman et al., 1999);
**$^{87}$Y** is a $\gamma$-emitting isotope that can be used to quantify the biodistribution of pharmaceuticals;
**$^{88}$Y** is a $\gamma$-emitting isotope used for calibration source
**$^{86}$Y** is a positron emitting isotope that can be used to quantify regional kinetics of yttrium via positron emission tomography (PET).
**$^{87m}$Sr** can be used in both diagnostic and therapeutic techniques for various skeletal diseases
**$^{83}$Sr** is $\beta^+$ emitter, potentially useful radionuclide for therapy planning prior to the use of the longer lived $\beta^+$ emitter $^{89}$Sr.
**$^{82}$Sr** is used exclusively to manufacture $^{82}$Sr/$^{82}$Rb generators widely used in PET myocardial perfusion imaging
**$^{85}$Sr** is a $\gamma$-emitting isotope used for calibration source
**$^{86}$Rb** is used as tracer in isotopic measurements of field metabolic rate (Bradshaw et al., 2007)
**$^{84}$Rb** is used in real-time tumor tracking system using implanted positron emission markers. Its half-live is comparable to the duration of a standard radiation therapy procedure.
**$^{82m}$Rb** was proposed as a substitute for cardio-PET clinical studies due to its suitable physical properties (Rowshanfarzad et al., 2006)
**$^{81}$Rb** is used through the $^{81}$Rb/$^{81m}$Kr generator where the short-lived eluent is used for investigation of pulmonary ventilation, or before surgical investigation.

In the frame of a systematic study of production routes of diagnostic and therapeutic radioisotopes we already made a series of measurements on excitation functions for the above mentioned activation products via proton, deuteron, alpha and $^3$He- particle induced reactions on krypton, rubidium, strontium, yttrium and zirconium targets (Tarkanyi et al., 1988a; Tarkanyi et al., 1988b; Qaim et al., 1988; Tarkanyi et al., 1990; Kovacs et al., 1991; Kovacs et al., 1991b; Fenyvesi et al.,1992; Kovacs et al., 1992; Qaim et al.,1994; Blessing et al., 1997; Ido et al., 2002a; Ido et al., 2002b; Tarkanyi et al., 2004 ; Uddin et al., 2005; Tarkanyi et al., 2005; Spahn et al., 2006; Uddin et al., 2006, ; Uddin et al., 2007; Tarkanyi et al., 2015). In these works some comparisons of the different production routes are also discussed but recently, new experimental data and reviews were published (Qaim et al., 2001; Hermanne et al., 2001; Filosofov et al., 2010; Betak et al., 2011; Zaneb et al., 2016; Capote et al., 2016) and we hence compare only shortly the production routes of the most widely used isotopes.

## Experimental procedure

The cross-section measurements were performed using the activation method combining a stacked foil irradiation and high resolution HPGe gamma-ray spectrometry. Cross-section data were deduced relative to the re-measured excitation functions of monitor reactions. The stack was irradiated in a short Faraday cup at an external beam line of the Cyclone 90 cyclotron of the Université Catholique in Louvain la Neuve (LLN) for 40 min. with a 50 MeV, 50 nA deuteron beam.

The stack contained a sequence of 10 blocks of Al (10 μm), $Sr(NO_3)_2$ (9-55 μm), Al (50 μm), Er (32 μm), Al (10 μm), $Ba(NO_3)_2$ (2 μm, sedimented), Al (50 μm) and Ag (10 μm) foils followed by 15 blocks of Ti (10 μm), Al (10 μm), $Ba(NO_3)_2$ (2 μm, sedimented), Al (50 μm) Ti (10 μm), Al (10 μm), $Ba(NO_2)$ (2 μm, sedimented), Al (50 μm). The 25 $Ba(NO_3)_2$ targets covered the 49.4-9.2 MeV energy range. The thin foils were purchased from Goodfellow Ltd., the compounds used for sedimentation were spectroscopic grade chemical materials from our storage. The sedimented targets were prepared in such a way that for the calculated mass of compound powder a given amount of evopreme glue was dissolved in chloroform and the compound powder was added to this solution under continuous ultrasonic mixing. The solution was poured in a 11 mm spot with Al backing and incubated until the chloroform content was completely evaporated and the compound was sedimented. The thicknesses of the targets were determined by weight and area measurement, in the case of sedimented samples the stoichiometry was also taken into account.

Four series of gamma-ray spectra were measured to follow the decay and were started at different times after end of bombardment (EOB): 3.9-7.1 h, 25.7-45.6 h, 92.1- 172.4 h and 1846.2- 1282.6 h, respectively. Due to the time required for the transfer of the irradiated targets from LLN to VUB (where the gamma spectra measurements were done), the large number of target foils irradiated in our irradiation session (4 stacks) and the limited detector capacity, it was impossible to optimize the gamma spectra measurements for all targets regarding all produced radioisotopes.

Gamma spectra were evaluated by the automatic fitting algorithm included in the Genie 2000 package (Genie) or in an iterative process using the Forgamma (Szekely, 1985) codes.

The decay data were taken from NUDAT2.6 (NUDAT2.6) and are shown in Table 1 together with the Q values of the contributing reactions (Qcalc). The preliminary data of the primary beam energy were taken from the settings of the accelerator; the beam intensity was derived from the Faraday cup results. The recommended data for the simultaneously re-measured excitation functions of the $^{27}Al(d,x)^{22,24}Na$ and $^{nat}Ti(d,x)^{48}V$ monitor reactions were taken from the IAEA database (Tarkanyi et al, 2001). The cross sections were determined using the well-known activation formula. Production cross sections on $^{nat}Sr$ were determined, not taking into account the abundance of the different naturally occurring Sr isotopes ($^{84}Sr$(0.56 %), $^{86}Sr$ (9.86 %), $^{87}Sr$ (7.00 %), $^{88}Sr$ (82.58 %).

Uncertainty on cross-sections was determined according to the recommendation in the ISO guide (Guide ISO) by taking the sum in quadrature of all individual linear contributions: beam current (7%), beam-loss corrections (max. 1.5%), target thickness (1%), detector efficiency (5%), photo peak area determination and counting statistics (1-20 %). The overall uncertainty hence ranges from 9 to 25%.

The median energies in targets were initially (for stack preparation) obtained from a degradation calculation based on the incident energy and the Andersen (Andersen and Ziegler, 1977) polynomial approximation of stopping powers. The complete excitation functions of the $^{27}Al(d,x)^{22,24}Na$ (Al from cover and backing) and $^{nat}Ti(d,x)^{48}V$ monitor reactions were re-measured simultaneously with the excitation functions of reactions induced on Ba targets to get final values for the beam energy and intensity in the Sr targets. If needed a corrections were applied based on the results of the fitted monitor reactions (final energy and intensity) (Tarkanyi et al., 1991). Uncertainty of energy was estimated by following the cumulative effects during the energy degradation of possible contributing uncertainties (primary energy, target thickness, energy straggling, correction for monitor reaction).

**Table 1** Decay characteristics of the investigated reaction products

| Nuclide<br>$J^{\pi}$<br>E-level<br>Decay mode | Half-life | $E_{\gamma}$ (keV) | $I_{\gamma}$(%) | Contributing process | Q-value (keV) |
|---|---|---|---|---|---|
| $^{88}$Y<br>4-<br>$\beta^+$ 100% | 106.627 d | 898.0<br>1836.1 | 93.7<br>99.2 | $^{87}$Sr(d,n)<br>$^{88}$Sr(d,2n) | 4483.1<br>-6629.5 |
| $^{87m}$Y<br>9/2+<br>380.82 keV<br>ε: 1.57 %<br>IT:98.43 % | 13.37 h | 380.8 | 78.1 | $^{86}$Sr(d,n)<br>$^{87}$Sr(d,2n)<br>$^{88}$Sr(d,3n) | 3559.5<br>-4868.6<br>-15981.3 |
| $^{87g}$Y<br>1/2-<br>ε: 100 %<br>$\beta^+$: 0.180 % | 79.8 h | 388.5<br>484.8 | 82.2<br>89.8 | $^{86}$Sr(d,n)<br>$^{87}$Sr(d,2n)<br>$^{88}$Sr(d,3n) | 3559.5<br>-4868.6<br>-15981.3 |
| $^{86g}$Y<br>4-<br>14.74 h<br>ε: 100 %<br>$\beta^+$ 31.9 % | 14.74 h | 443.1<br>627.7<br>703.3<br>777.4<br>1076.6<br>1153.1 | 16.9<br>32.6<br>15.4<br>22.4<br>82.5<br>30.5 | $^{86}$Sr(d,2n)<br>$^{87}$Sr(d,3n)<br>$^{88}$Sr(d,4n) | -8246.9<br>-16675.1<br>-27787.7 |
| $^{85g}$Y<br>(1/2)-<br>ε: 100 %<br>$\beta^+$ 66 % | 2.68 h | 231.65<br>504.44 | 84<br>60 | $^{84}$Sr(d,n)<br>$^{86}$Sr(d,3n)<br>$^{87}$Sr(d,4n)<br>$^{88}$Sr(d,5n) | 2257.0<br>-17759.3<br>-26187.5<br>-37300.1 |
| $^{87m}$Sr<br>$1/2^-$<br>388.53 keV<br>ε : 0.3 %<br>IT : 99.7 % | 2.815 h | 388.5 | 82.2 | $^{86}$Sr(d,p)<br>$^{87}$Sr(d,pn)<br>$^{88}$Sr(d,p2n) | 6203.6<br>-2224.6<br>-13337.2 |
| $^{85m}$Sr<br>1/2-<br>238.795 keV<br>ε: 13.4 %<br>IT : 86.6% | 67.63 min | 151.2<br>231.9 | 12.8<br>83.9 | $^{84}$Sr(d,p)<br>$^{86}$Sr(d,p2n)<br>$^{87}$Sr(d,p3n)<br>$^{88}$Sr(d,p4n)<br>$^{85}$Y decay | 6300.47<br>-13715.81<br>-22144.0<br>-33256.6<br>-2257.0 |
| $^{83}$Sr<br>7/2+<br>ε: 100 %<br>$\beta^+$ 27 % | 32.41 h | 381.5<br>762.7 | 14.0<br>26.7 | $^{84}$Sr(d,p2n)<br>$^{86}$Sr(d,p4n)<br>$^{87}$Sr(d,p5n)<br>$^{88}$Sr(d,p6n)<br>$^{83}$Y decay | -14147.8<br>-34164.1<br>-42592.3<br>-53704.9<br>-19523.3 |
| $^{82}$Sr<br>0+<br>ε: 100 % | 25.35 d | | | $^{84}$Sr(d,p3n)<br>$^{86}$Sr(d,p5n)<br>$^{87}$Sr(d,p6n)<br>$^{88}$Sr(d,p7n)<br>$^{82}$Y decay | -23006.7<br>-43023.0<br>-51451.2<br>-62563.8<br>-31736.2 |
| $^{82}$Rb<br>1+<br>ε: 100 %<br>$\beta^+$ :95.43 % | 1.2575 min | 776.5 | 15.1 | $^{82}$Sr decay | -23006.7 |
| $^{86}$Rb<br>2-<br>ε: 0.0052<br>$\beta^-$ 100% | 18.642 d | 1077.0 | 8.6 | $^{86}$Sr(d,2p)<br>$^{87}$Sr(d,2pn)<br>$^{88}$Sr(d,2p2n) | -3218.4<br>-11646.6<br>-22759.2 |
| $^{84}$Rb | 32.82 d | 881.6 | 68.9 | $^{84}$Sr(d,2p) | -2332.8 |
| 2-<br>ε: 96.1%<br>$\beta^-$: 3.9 % | | | | $^{86}$Sr(d,2p2n)<br>$^{87}$Sr(d,2p3n)<br>$^{88}$Sr(d,2p4n) | -22349.1<br>-30777.3<br>-42062.9 |
| $^{83}$Rb<br>5/2-<br>ε: 100 % | 86.2 d | 520.4<br>529.6<br>552.6 | 45<br>29.3<br>16.0 | $^{84}$Sr(d,2pn)<br>$^{86}$Sr(d,2p3n)<br>$^{87}$Sr(d,2p4n)<br>$^{88}$Sr(d,2p5n)<br>$^{83}$Sr decay | -11092.6<br>-31108.7<br>-39536.9<br>-50649.5<br>-31736.2 |
| $^{82m}$Rb<br>5-<br>68.9*1 keV*<br>ε: 1001%<br>$\beta^+$ 21.2 % | 6.472 h | 554.4<br>619.1<br>698.4<br>776.5<br>827.8<br>1044.1<br>1317.4<br>1474.9 | 62.4<br>38.0<br>26.3<br>84.4<br>21.0<br>32.1<br>23.7<br>15.5 | $^{84}$Sr(d,2p2n)<br>$^{86}$Sr(d,2p4n)<br>$^{87}$Sr(d,2p5n)<br>$^{88}$Sr(d,2p6n) | -22046.6<br>-42062.9<br>-50491.1<br>-62217.9 |
| $^{81}$Rb<br>3/2-<br>ε: 100 %<br>$\beta^+$ 27.2 % | 4.572 h | 190.5<br>446.2 | 64.9<br>23.5 | $^{84}$Sr(d,2p3n)<br>$^{86}$Sr(d,2p5n)<br>$^{87}$Sr(d,2p6n)<br>$^{88}$Sr(d,2p7n)<br>$^{81}$Sr decay | -30849.1<br>-50865.4<br>-59293.5<br>-70406.1<br>-35559.9 |

Abundances in $^{nat}$Sr(%): $^{84}$Sr(0.56), $^{86}$Sr (9.86), $^{87}$Sr (7.00), $^{88}$Sr (82.58)

Increase Q-values if compound particles are emitted: np-d, +2.2 MeV; 2np-t, +8.48 MeV; n2p-$^{3}$He, +7.72 MeV; 2n2p-$\alpha$, +28.30 MeV.

Decrease Q-values for isomeric states with level energy of the isomer

## Results

The measured cross-sections for the production of $^{88,87m,87g,86g,85g}Y$, $^{87m,85m,83g,82}Sr$, $^{86g,84g,83,82m,81}Rb$ are shown in Tables 2-4 and Figures 1-15. The figures also show the theoretical results available in the TALYS 1.6 (Koning et al., 2012) based TENDL-2015 library (Koning et al., 2015). We are presenting cumulative results in several cases, when the half-life of the isomeric state decaying into the ground state is much sorter then that of the ground state. In this case we use the results of a late measurement, when the isomeric state is completely decayed out. The TENDL results are constructed in this case by adding the direct isomeric cross section to the direct ground state cross section.

Table 2. Activation cross-sections for formation of $^{88,87m,87g,86g,85m}Y$ in $^{nat}Sr(d,xn)$ reactions.

| | | $^{88}Y$ | | $^{87m}Y$ | | $^{87g}Y$ | | $^{86g}Y$ | | $^{85m}Y$ | |
|---|---|---|---|---|---|---|---|---|---|---|---|
| **E** | **ΔE** | **σ** | **Δσ** | **σ** | **Δσ** | **σ** | **Δσ** | **σ** | **Δσ** | **σ** | **Δσ** |
| MeV | | mb | | | | | | | | | |
| 49.95 | 0.20 | 94.9 | 11.6 | 234.1 | 26.4 | 358.0 | 40.2 | 427.2 | 48.5 | | |
| 48.06 | 0.25 | 90.2 | 12.9 | 209.2 | 23.7 | 346.6 | 38.9 | 370.2 | 42.1 | 36.6 | 18.6 |
| 46.13 | 0.30 | 122.9 | 15.2 | 284.3 | 32.2 | 443.5 | 49.9 | 464.4 | 54.3 | | |
| 44.14 | 0.35 | 135.2 | 17.9 | 283.1 | 31.9 | 473.8 | 53.2 | 435.1 | 49.3 | | |
| 42.11 | 0.40 | 106.5 | 12.8 | 281.6 | 31.7 | 426.5 | 47.9 | 362.8 | 41.7 | 33.8 | 13.4 |
| 40.01 | 0.45 | 141.1 | 16.7 | 370.9 | 41.7 | 550.3 | 61.8 | 393.7 | 44.6 | 33.9 | 16.2 |
| 37.85 | 0.50 | 179.4 | 23.9 | 432.9 | 48.8 | 694.1 | 77.9 | 316.7 | 36.6 | | |
| 35.62 | 0.56 | 179.4 | 21.0 | 544.9 | 61.3 | 804.1 | 90.3 | 269.1 | 30.9 | | |
| 33.32 | 0.62 | 233.9 | 28.9 | 671.8 | 75.6 | 1052.4 | 118.2 | 173.8 | 23.6 | | |
| 30.92 | 0.68 | 239.8 | 27.9 | 741.7 | 83.4 | 1090.8 | 122.5 | 114.5 | 15.9 | 18.1 | 17.1 |
| 28.42 | 0.74 | 289.3 | 33.5 | 764.7 | 84.1 | 1125.3 | 126.4 | 109.9 | 14.3 | | |
| 24.80 | 0.83 | 414.9 | 47.2 | 657.4 | 73.9 | 1024.8 | 115.1 | 117.9 | 15.4 | | |
| 22.60 | 0.88 | 509.5 | 58.2 | 475.9 | 53.5 | 792.8 | 89.0 | 88.7 | 13.2 | | |
| 20.29 | 0.94 | 685.8 | 77.4 | 336.5 | 37.9 | 535.1 | 60.1 | 114.9 | 14.7 | | |
| 18.33 | 0.99 | 819.0 | 92.1 | 142.6 | 16.1 | 222.2 | 24.9 | 98.9 | 11.3 | | |
| 16.78 | 1.03 | 871.1 | 98.0 | 92.7 | 10.5 | 147.5 | 16.6 | 91.2 | 11.0 | | |
| 15.17 | 1.07 | 823.1 | 92.6 | 77.2 | 8.7 | 121.7 | 13.7 | 80.5 | 9.4 | | |
| 13.48 | 1.11 | 612.8 | 69.0 | 60.5 | 6.9 | 109.3 | 12.3 | 41.6 | 5.2 | | |
| 11.71 | 1.16 | 236.5 | 26.8 | 40.7 | 4.6 | 76.6 | 8.6 | 10.5 | 1.7 | | |
| 9.83 | 1.20 | 10.1 | 1.4 | 4.0 | 0.5 | 8.4 | 1.0 | | | | |
| 7.81 | 1.25 | 34.5 | 4.5 | 4.6 | 0.7 | 8.5 | 1.0 | | | | |
| 5.64 | 1.31 | | | 1.4 | 0.2 | | | | | | |
| 3.24 | 1.37 | 6.4 | 1.2 | 0.9 | 0.2 | 2.1 | 0.3 | | | | |
| 0.54 | 1.44 | 2.5 | 0.4 | 0.4 | 0.1 | 0.8 | 0.1 | | | | |

Table 3. Activation cross-sections for formation of $^{85g}$Y, $^{87m,85g,83g,82}$Sr in $^{nat}$Sr(d,x) reactions

| | | $^{85g}$Y | | $^{87m}$Sr | | $^{85g}$Sr | | $^{83g}$Sr | | $^{82}$Sr | |
|---|---|---|---|---|---|---|---|---|---|---|---|
| E | ΔE | σ | Δσ | σ | Δσ | σ | Δσ | σ | Δσ | σ | Δσ |
| MeV | | mb | | | | | | | | | |
| 49.95 | 0.20 | 29.4 | 3.4 | 50.0 | 9.1 | 260.9 | 29.3 | 7.4 | 3.1 | | |
| 48.06 | 0.25 | 19.3 | 2.3 | 42.4 | 8.0 | 187.7 | 21.1 | 9.2 | 1.5 | 2.7 | 1.1 |
| 46.13 | 0.30 | 16.1 | 2.1 | 52.3 | 9.7 | 192.5 | 21.7 | | | 2.8 | 1.8 |
| 44.14 | 0.35 | 16.6 | 2.0 | 44.5 | 8.7 | 163.0 | 18.3 | 3.4 | 1.6 | 2.3 | 1.3 |
| 42.11 | 0.40 | 14.2 | 1.8 | 40.3 | 8.5 | 127.0 | 14.3 | | | | |
| 40.01 | 0.45 | 15.7 | 2.0 | 42.9 | 10.0 | 141.7 | 15.9 | 2.8 | 0.3 | | |
| 37.85 | 0.50 | 16.9 | 2.2 | 34.4 | 9.4 | 155.6 | 17.5 | | | | |
| 35.62 | 0.56 | 21.8 | 2.8 | 37.5 | 12.7 | 146.7 | 16.7 | 7.3 | 4.1 | | |
| 33.32 | 0.62 | 18.0 | 2.3 | 35.4 | 10.9 | 140.7 | 15.9 | 6.5 | 2.3 | | |
| 30.92 | 0.68 | 18.5 | 2.4 | 33.6 | 12.4 | 139.0 | 15.7 | | | | |
| 28.42 | 0.74 | 20.4 | 2.4 | 25.2 | 9.3 | 122.4 | 13.8 | | | | |
| 24.80 | 0.83 | 14.6 | 1.9 | 12.6 | 9.8 | 89.3 | 8.0 | | | | |
| 22.60 | 0.88 | 9.0 | 1.2 | 11.4 | 6.6 | 52.7 | 5.9 | | | | |
| 20.29 | 0.94 | 3.7 | 0.7 | 12.2 | 5.5 | 17.7 | 2.0 | | | | |
| 18.33 | 0.99 | | | 13.1 | 3.9 | 8.1 | 0.9 | | | | |
| 16.78 | 1.03 | | | 10.9 | 3.1 | 6.7 | 0.8 | | | | |
| 15.17 | 1.07 | | | 8.3 | 2.61 | 5.3 | 0.6 | | | | |
| 13.48 | 1.11 | | | 7.5 | 2.8 | 4.9 | 0.6 | | | | |
| 11.71 | 1.16 | 0.7 | 0.1 | 6.3 | 2.0 | 4.3 | 0.5 | | | | |
| 9.83 | 1.20 | | | 2.4 | 0.5 | 1.2 | 0.2 | | | | |
| 7.81 | 1.25 | | | 1.8 | 0.5 | | | | | | |
| 5.64 | 1.31 | | | 1.1 | 0.3 | | | | | | |
| 3.24 | 1.37 | | | 0.6 | 0.2 | | | | | | |
| 0.54 | 1.44 | | | | | | | | | | |

Table 4. Activation cross-sections for formation of $^{86g,84g,83,82m,81g}Rb$ in $^{nat}Sr(d,x)$ reactions

| | | $^{86g}$Rb | | $^{84g}$Rb | | $^{83}$Rb | | $^{82m}$Rb | | $^{81g}$Rb | |
|---|---|---|---|---|---|---|---|---|---|---|---|
| **E** | **Δσ** | σ | **Δσ** | σ | **Δσ** | σ | **Δσ** | σ | **Δσ** | σ | **Δσ** |
| MeV | | mb | | | | | | | | | |
| 49.95 | 0.20 | 11.7 | 3.8 | 33.3 | 4.1 | 76.9 | 9.4 | 11.7 | 2.1 | 13.4 | 1.6 |
| 48.06 | 0.25 | | | 29.3 | 3.5 | 70.2 | 9.2 | 15.3 | 2.4 | 11.2 | 1.4 |
| 46.13 | 0.30 | 11.2 | 5.3 | 43.6 | 5.2 | 65.8 | 8.5 | 14.5 | 2.7 | 12.3 | 1.6 |
| 44.14 | 0.35 | | | 44.8 | 5.3 | 54.6 | 7.7 | 11.6 | 2.1 | 9.7 | 1.3 |
| 42.11 | 0.40 | | | 39.1 | 4.5 | 31.7 | 4.3 | 6.2 | 1.4 | 6.9 | 0.9 |
| 40.01 | 0.45 | | | 46.7 | 5.6 | 21.4 | 3.4 | 9.0 | 1.8 | 6.0 | 1.5 |
| 37.85 | 0.50 | | | 49.1 | 5.9 | 33.1 | 6.1 | 5.2 | 1.8 | 2.7 | 1.3 |
| 35.62 | 0.56 | 6.9 | 3.1 | 48.6 | 5.6 | 23.2 | 3.5 | | | | |
| 33.32 | 0.62 | | | 38.9 | 5.1 | 23.8 | 5.4 | 3.7 | 1.5 | | |
| 30.92 | 0.68 | | | 36.8 | 4.6 | 17.5 | 3.2 | | | | |
| 28.42 | 0.74 | | | 24.5 | 3.1 | 13.6 | 3.0 | | | | |
| 24.80 | 0.83 | | | 7.3 | 2.3 | 13.6 | 2.6 | | | | |
| 22.60 | 0.88 | | | 5.6 | 1.3 | | | | | | |
| 20.29 | 0.94 | | | 4.6 | 1.1 | | | | | | |
| 18.33 | 0.99 | | | 1.9 | 0.4 | | | | | | |
| 16.78 | 1.03 | | | 1.4 | 0.4 | | | | | | |
| 15.17 | 1.07 | | | 1.3 | 0.3 | | | | | | |
| 13.48 | 1.11 | | | | | | | | | | |
| 11.71 | 1.16 | | | | | | | | | | |
| 9.83 | 1.20 | | | | | | | | | | |
| 7.81 | 1.25 | | | | | | | | | | |
| 5.64 | 1.31 | | | | | | | | | | |
| 3.24 | 1.37 | | | | | | | | | | |
| 0.54 | 1.44 | | | | | | | | | | |

### The $^{nat}Sr(d,x)^{88}Y$ reaction

The excitation function for production of $^{88}Y$ ($T_{1/2}$ = 106.627 d) is shown in Fig. 1. The agreement in magnitude with the theory is acceptable, but the threshold is shifted.

### The $^{nat}Sr(d,x)^{87m}Y$ reaction

The independent cross sections for formation of the metastable state of the $^{87m}Y$ ($T_{1/2}$ = 13.37 h) are shown in Fig. 2 in comparison with TENDL-2015. The agreement with the theory is good, except between 5 and 13 MeV, where the experiment does not show so large low energy contribution.

### The $^{nat}Sr(d,x)^{87g}Y$ reaction.

The excitation function for production of the ground state of $^{87}Y$ ($T_{1/2}$ = 79.8 h) was measured after complete decay of isomeric state ($T_{1/2}$ = 13.37 h, IT:98.43 %) and is hence cumulative. The theory underestimates the experimental data (Fig. 3).

### The $^{nat}Sr(d,x)^{86g}Y$ reaction

The cross sections for production of ground state of $^{86}Y$ ($T_{1/2}$ = 14.74 h) were obtained after the complete decay of shorter lived isomeric state ($T_{1/2}$= 47.4 min). The description by TENDL-2015 is good (Fig. 4).

### The $^{nat}Sr(d,xn)^{85m}Y$ reaction

The radionuclide $^{85}Y$ has two isomeric states: the shorter-lived higher lying state ($T_{1/2}$ = 4.86 h) and the ground state ($T_{1/2}$ = 2.68 h). They are decaying independently. We could obtain independent cross- section data for production of both states. Only a few cross-section points could be derived for the isomeric state. According to Fig. 5 the theory overestimates the experiment.

### The $^{nat}Sr(d,xn)^{85g}Y$ reaction

The cross-sections for the $^{85g}Y$ ground state ($T_{1/2}$ = 2.68 h) were measured with acceptable statistics at all energy points (Fig. 6). The theory underestimates the experiment.

### The $^{nat}Sr(d,x)$ $^{87m}Sr$ reaction

The cross-sections for $^{87m}Sr$ ($T_{1/2}$ = 2.815 h) are shown in Fig. 7. Some shift in the shape can be observed in comparison with the TENDL-2015 results. The metastable $^{87m}Sr$ is produced directly and through the decay of $^{87m}Y$ (13.37 h, ε: 1.57 %) and $^{87g}Y$(79.8 h, ε: 100 %) parents. These indirect production paths were not taken into account in the presented TENDL curve, the good agreement means that the contribution of them is very low.

### The $^{nat}Sr(d,x)^{85g}Sr$ reaction

The cross-section for $^{85g}Sr$ ($T_{1/2}$ = 64.849 d) are cumulative, formation is direct and through the decay of the shorter-lived isomeric state ($T_{1/2}$ = 67.63 min) and through the decay of the isomeric states of the parent isotope $^{85m}Y$ ($T_{1/2}$ = 4.86 h) and $^{85g}Y$ ($T_{1/2}$ = 2.68 h) (Fig. 8). Deviation between experiment and theory (without indirect production paths) occurs above 40 MeV.

### The $^{nat}Sr(d,x)^{83g}Sr$ reaction

The measured cumulative cross section of the $^{83g}Sr$ ($T_{1/2}$ = 32.41 h) ( Fig. 9) contains the decay contribution of the $^{83m}Sr$ ($T_{1/2}$ = 4.95 s, IT: 100 %) and of the parent $^{83}Y$ ($T_{1/2}$ = 2.85 min**)**. Only a few experimental data were obtained, which are in agreement with the theory.

### The $^{nat}Sr(d,x)^{82}Sr$ reaction

No gamma lines are emitted in the decay of $^{82}Sr$ ($T_{1/2}$ = 25.35 d) therefore the production cross sections has to be assessed through the decay of its shorter-lived daughter $^{82m}Rb$( $T_{1/2}$ = 1.2575 min). Only three data points were obtained (Fig. 10) fitting to the theoretical prediction. The $^{82}$Sr is produced directly and through the decay of $^{82}$Y ($T_{1/2}$ = 9.5 s).

### The $^{nat}$Sr(d,x)$^{86g}$Rb reaction

The production cross-sections of the ground state of the closed nucleus $^{86}$Rb ($T_{1/2}$ = 18.642 d) include the contributions of the short-lived $^{86m}$Rb isomeric state ($T_{1/2}$ = 1.017 min, IT 100 %). The obtained few cross-sections are in good agreement with the theory (Fig. 11).

### The $^{nat}$Sr(d,x)$^{84g}$Rb reaction

The $^{84}$Rb is also a closed nucleus. Here cumulative production cross-sections for the ground state ($T_{1/2}$ = 32.82 d) after complete decay of the metastable state ($T_{1/2}$ = 20.26 min, IT= 100%) are presented (Fig. 12). The experimental values are a multiple of the TENDL-2015 predictions above 20 MeV.

### The $^{nat}$Sr(d,x)$^{83g}$Rb reaction

The excitation function for cumulative production of $^{83}$Rb ($T_{1/2}$ = 86.2 d), including full decay of the short lived isomeric states of $^{83}$Y ($m_1$:$T_{1/2}$ = 2.85 min, ε: 60 *5* %, IT: 40 *5%* and $m_2$ :$T_{1/2}$ = 7.1 min, ε: 100 %) and $^{83}$Sr ($m_1$: $T_{1/2}$ = 4.95 s and $m_2$: $T_{1/2}$ = 32.41 h) are shown in Fig. 13.

### The $^{nat}$Sr(d,x)$^{82m}$Rb reaction

The production cross sections for direct production of $^{82m}$Rb ($T_{1/2}$ = 6.472 h) is shown in Fig. 14 in comparison with the theory. The agreement is partly acceptable.

### The $^{nat}$Sr(d,x)$^{81}$Rb reaction

The excitation function for cumulative production of the ground state of $^{81}$Rb ($T_{1/2}$ = 4.572 h) after the complete decay of the short-lived $^{81m}$Rb isomeric state ($T_{1/2}$ = 30.5 min, IT = 97.6 6%) and the simultaneously produced $^{81}Y$ ($T_{1/2}$ = 70.4 s, ε: 100 %) and $^{81}Sr$ ($T_{1/2}$ = 22.3 min, ε: 100 %) parent nuclei is shown in Fig. 15.

## Integral yields

Based on the experimentally determined cross-sections we have calculated the differential and integral yields of all studied produced radioisotopes. The yields are so called physical yields, calculated for an instantaneous irradiation (Bonardi 1987, Otuka and Takacs, 2015) (Fig. 16 and 17).

The calculated yields for production of $^{86,87m,87,88}Y$ are compared with the experimental yield data measured by Dmitriev at 22 MeV, (Dmitriev et al , 1983) on Fig. 16 , showing large disagreements for $^{87m}Y$ and $^{87g}Y$ .

## Production $^{88,87,86}Y$ and $^{89}Sr$

We shortly discuss only the production of radioisotopes, which can be produced carrier free ($^{88,87,86}Y$) and for which the yields of deuteron induced reactions are comparable to other production routes. We also include production of long-lived $^{89}Sr$, which can in principle be produced via a (d,p) reaction. Unfortunately, in spite of long spectra measurements, we could not get reliable cross section data with the used technology, due to the only available very weak gamma-line ($T_{1/2}$ = 50.563 d, $E_\gamma$ = 908.960 keV, $I_\gamma$ = 0.00956 %). Our effort was also unsuccessful in our earlier investigation through alpha induced reaction on krypton (Tarkanyi et al., 1990).

By analyzing the nuclear data from the point of view of the applications the following conclusions can be deduced for the deuteron induced reactions of the above mentioned activation products.

In all earlier practical productions of these radioisotopes neutron, proton and alpha particle induced reactions were used on Sr, Rb and Zr targets. The selection is based on well-known factors as the availability of accelerators, the production yield, the target availability, price and recovery problems, the carrier and radioisotopic impurity of the end-product, etc.

Some general conclusions exist:

When it is possible the proton induced reactions are preferred: availability of beam in all energy range from commercial accelerators, the available high beam intensity, smallest energy degradation, and relatively high cross sections.

Although deuterons also can be accelerated as negative ions at cyclotrons and can be extracted with high efficiency, the maximum energy is half of that of the protons and the energy degradation is higher. Beam intensities are lower, but the cross sections are often comparable to those of proton induced reaction. Especially the (d,2n) reaction for middle mass targets is more productive then the (p,n). Also the (d,n) and (d,p) reactions make such radioisotopes available that cannot be produced with protons.

The alpha induced reactions are only practical when proton or deuteron reactions are not allowing the aimed route for production (only a limited number of interesting radionuclides). A limited number of accelerators with alpha option exist and their number is decreasing. The α-reaction cross-sections are moderate, the beam intensity is low and the energy degradation is high.

The comparison of the excitation functions for direct production of $^{88,87,86}Y$ and $^{89}Sr$ by different incident particles are shown in Fig. 18-21. Cross-section data are based on TENDL-2015 library, compared and cross checked with the available literature experimental data.

Long-lived **$^{88}$Y** ($T_{1/2}$ = 106.627 d) can be produced directly through the $^{88}Sr$ (p,n), $^{88}Sr$ (d,2n), $^{85}Rb$(α,n), $^{87}Rb(^{3}He,2n)^{88}Y$ routes and through decay of its long half-life $^{88}Zr$ ($T_{1/2}$ = 83.4 d) parent isotope formed in following reactions: $^{89}Y(p,2n)^{88}Zr(^{88}Y)$, $^{89}Y(d,3n)^{88}Zr(^{88}Y)$, $^{nat}Zr(p,x)^{88}Zr(^{88}Y)$, $^{nat}Zr(d,x)^{88}Zr(^{88}Y)$.

Here we compare only the direct production routes (Fig. 18). The presently investigated $^{88}Sr$(d,2n) has a production yield comparable to the $^{88}Sr$(p,n) reaction, but it requires a higher energy cyclotron and additionally a small amount of contaminating stable $^{89}Y$ is produced. The (α,n) and ($^{3}He$,2n) have significantly lower yield and are of no practical importance.

The radionuclide **$^{87}$Y** ($T_{1/2}$ = 13.37 h) can be produced directly through the $^{88}Sr$(p,2n), $^{86}Sr$(d,n), $^{87}Sr$(d,2n), $^{85}Rb$(α,2n), $^{85}Rb(^{3}He,n)$ routes and through the decay of its shorter half-life $^{87}Zr$ ($T_{1/2}$ = 1.68 h) parent isotope obtained through the following reactions: $^{89}Y(p,3n)^{87}Zr$ ($^{87}Y$), $^{89}Y(d,4n)^{87}Zr(^{87}Y)$, $^{nat}Zr$(p,x), $^{nat}Zr$(d,x), $^{89}Y(p,3n)^{87}Zr$ ($^{87}Y$). Here we compare only the direct production routes (Fig. 19).

Only the $^{87}Sr$(p,n) and $^{86}Sr$(d,n) reactions are, where no long half-life $^{88}Y$ is simultaneously produced, but the yield of the (d,n) is significantly lower. If contamination is not so important, the $^{88}Sr$(p,2n) and $^{87}Sr$(d,2n) reactions are the most productive. The $^{88}Sr$(d,3n) also has high yield but requires higher energy machine. In spite of the interesting cross-sections the $^{85}Rb$(α,2n) reaction has lower yield due to the higher energy degradation.

The **$^{86}$Y** (T$_{1/2}$ = 14.74 h) can be produced directly through the $^{86}$Sr(p,n), $^{87}$Sr(p,2n) $^{86}$Sr(d,2n), $^{85}$Rb (α,3n), $^{85}$Rb($^{3}$He,2n) reactions and through decay of its shorter half-life parent isotope $^{86}$Zr (T$_{1/2}$ = 16.5 h) using the $^{nat}$Zr(p,x), $^{nat}$Zr(d,x), $^{89}$Y(p,4n)$^{86}$Zr ($^{86}$Y) reactions (Fig. 20).

The (p,n) production results in a high radionuclidic purity, carrier free product. In case of all other reactions $^{87}$Y is simultaneously produced and in case of (α,3n) also $^{88}$Y. The production yields are the highest for the proton and deuteron induced reactions.

The **$^{89}$Sr** can be produced carrier free via the widely used $^{89}$Y(n,p) reaction and by using $^{86}$Kr(α,n) (Fig. 21). In case of $^{88}$Sr(n,γ) and $^{88}$Sr(d,p) the product is carrier added.

## Summary and conclusion

We report experimental cross-sections for the nuclear reactions $^{nat}$Sr(d,x)$^{88,87m,87g,86g,85g}$Y, $^{87m,85g,83g,82}$Sr and $^{86g,84g,83,82m,81g}$Rb up to 50 MeV deuteron energies. Our study makes available the first experimental activation data sets for all products. The comparison with the TENDL-2015 library shows acceptable agreement in most cases, taking into account the well-known complexity of the description for deuterons.

The possible use of deuteron induced reactions for production of $^{88,87,86}$Y and $^{89}$Sr was discussed in comparison with other production routes. It is shown that for these selected radionuclides deuteron induced reactions are of practical importance.

## Figures

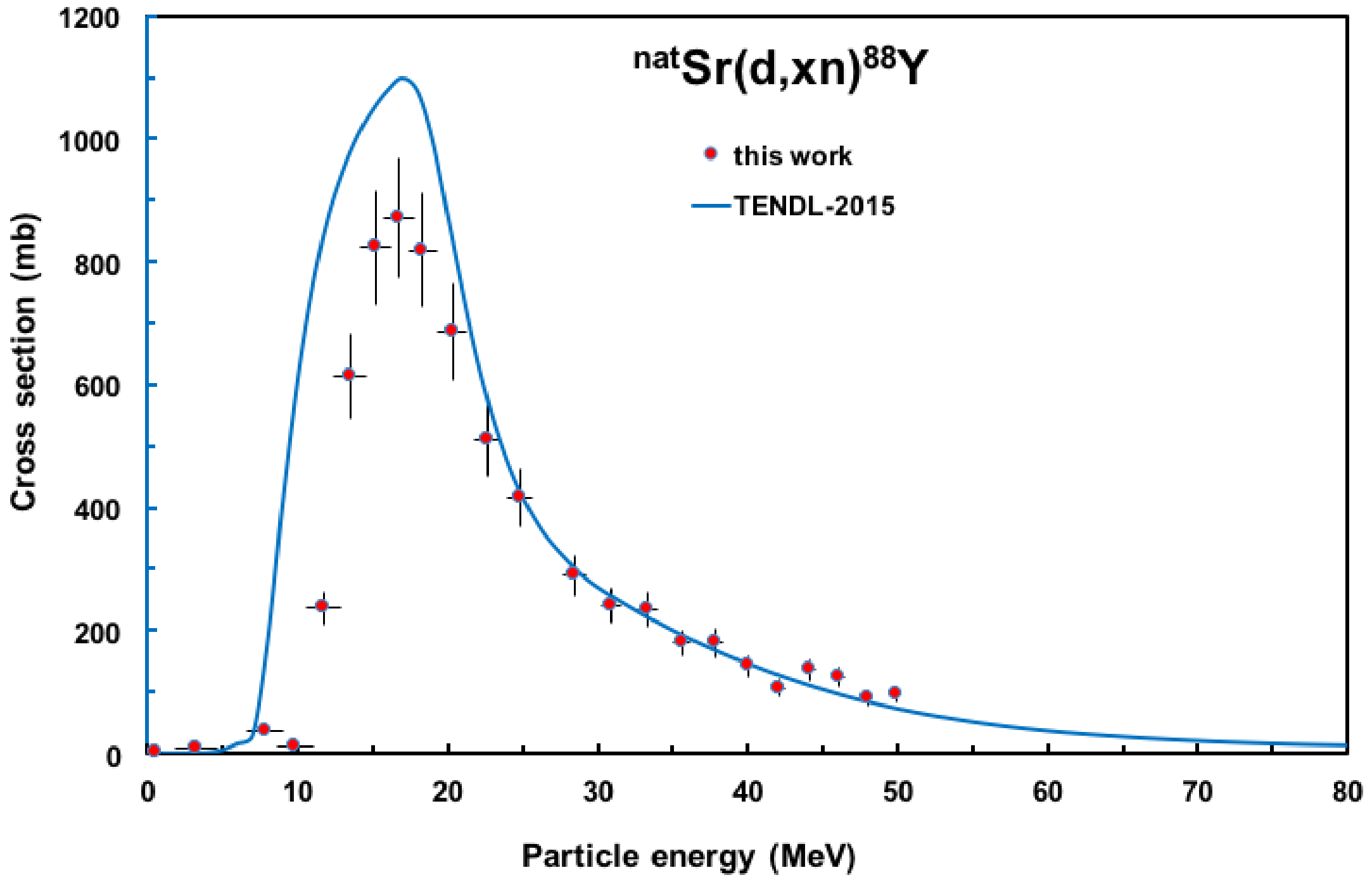

Fig. 1. Excitation function of the ${}^{nat}Sr(d,x){}^{88}Y$ reaction

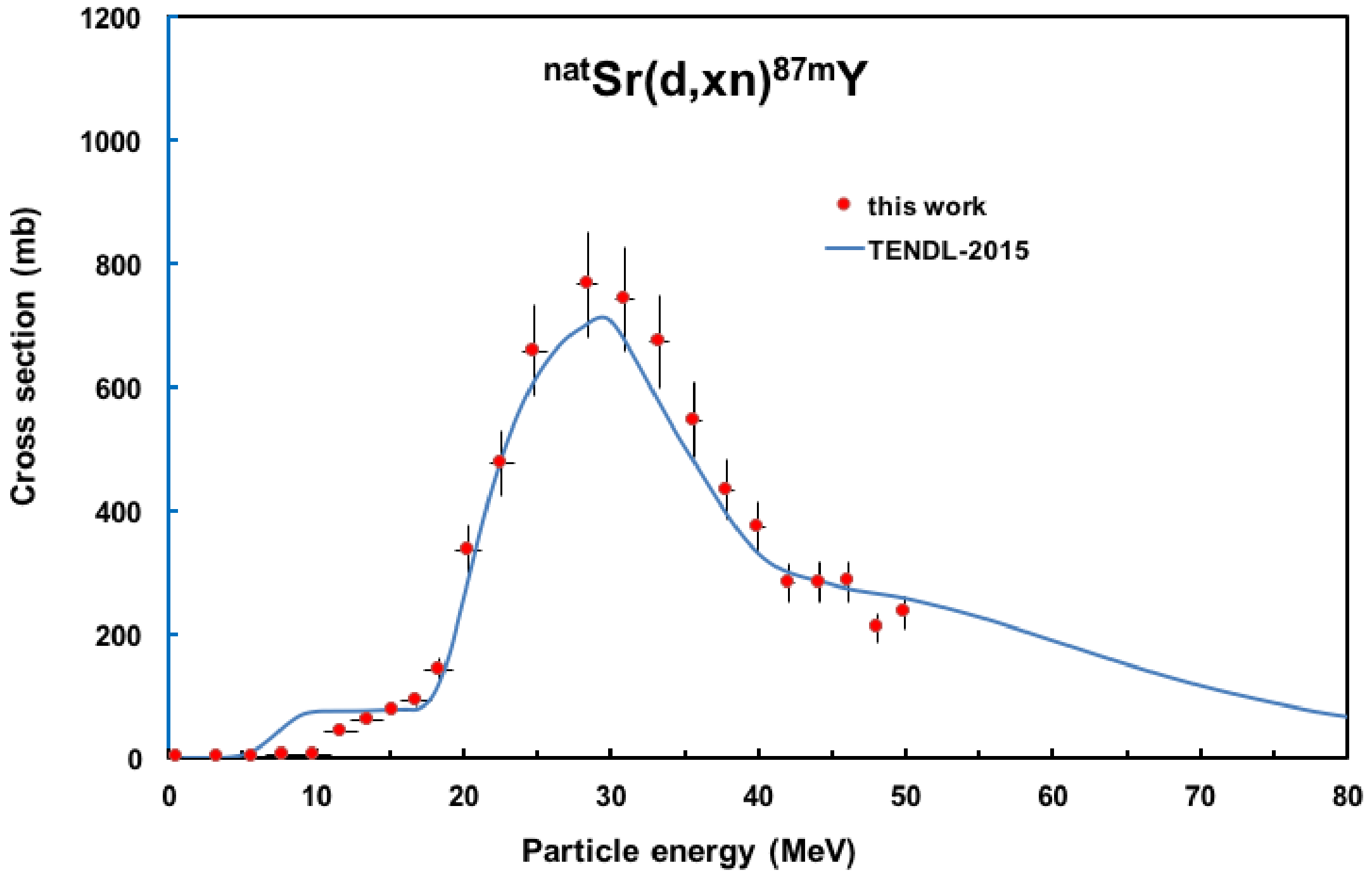

Fig. 2. Excitation function of the ${}^{nat}Sr(d,x){}^{87m}Y$ reaction

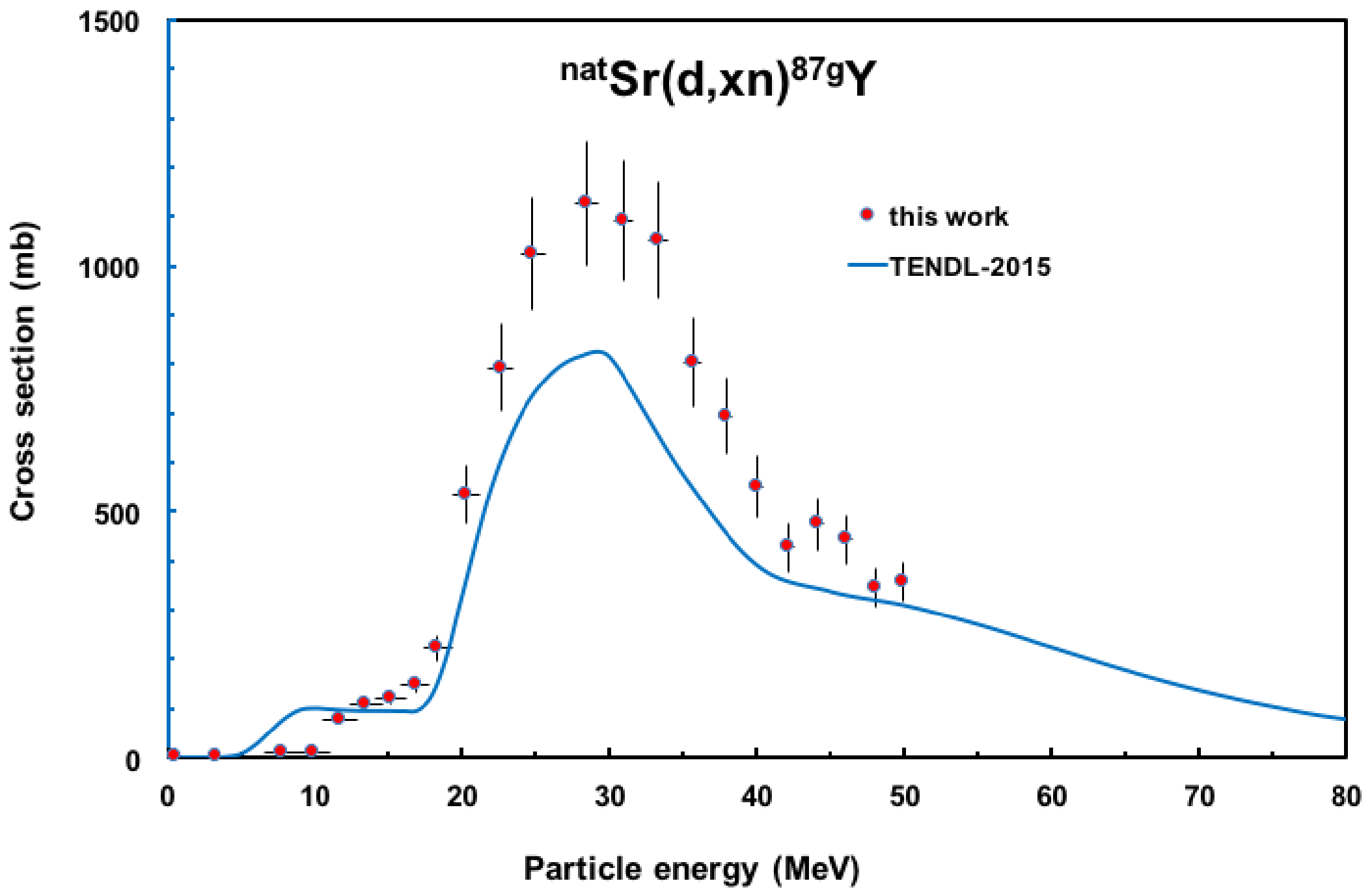

Fig. 3. Excitation function of the ${}^{nat}Sr(d,x)^{87g}Y$ reaction

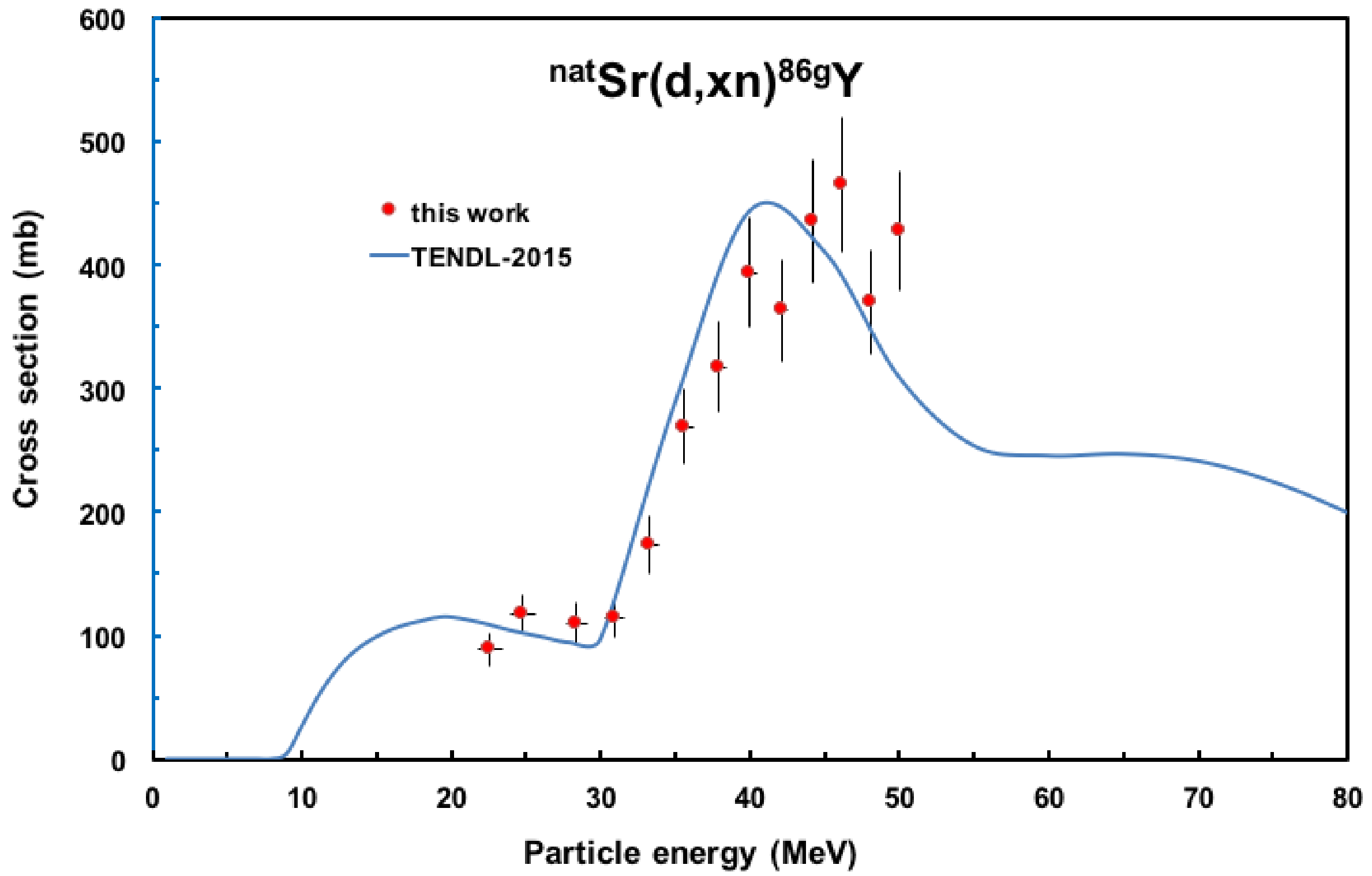

Fig. 4. Excitation function of the ${}^{nat}Sr(d,x)^{86g}Y$ reaction

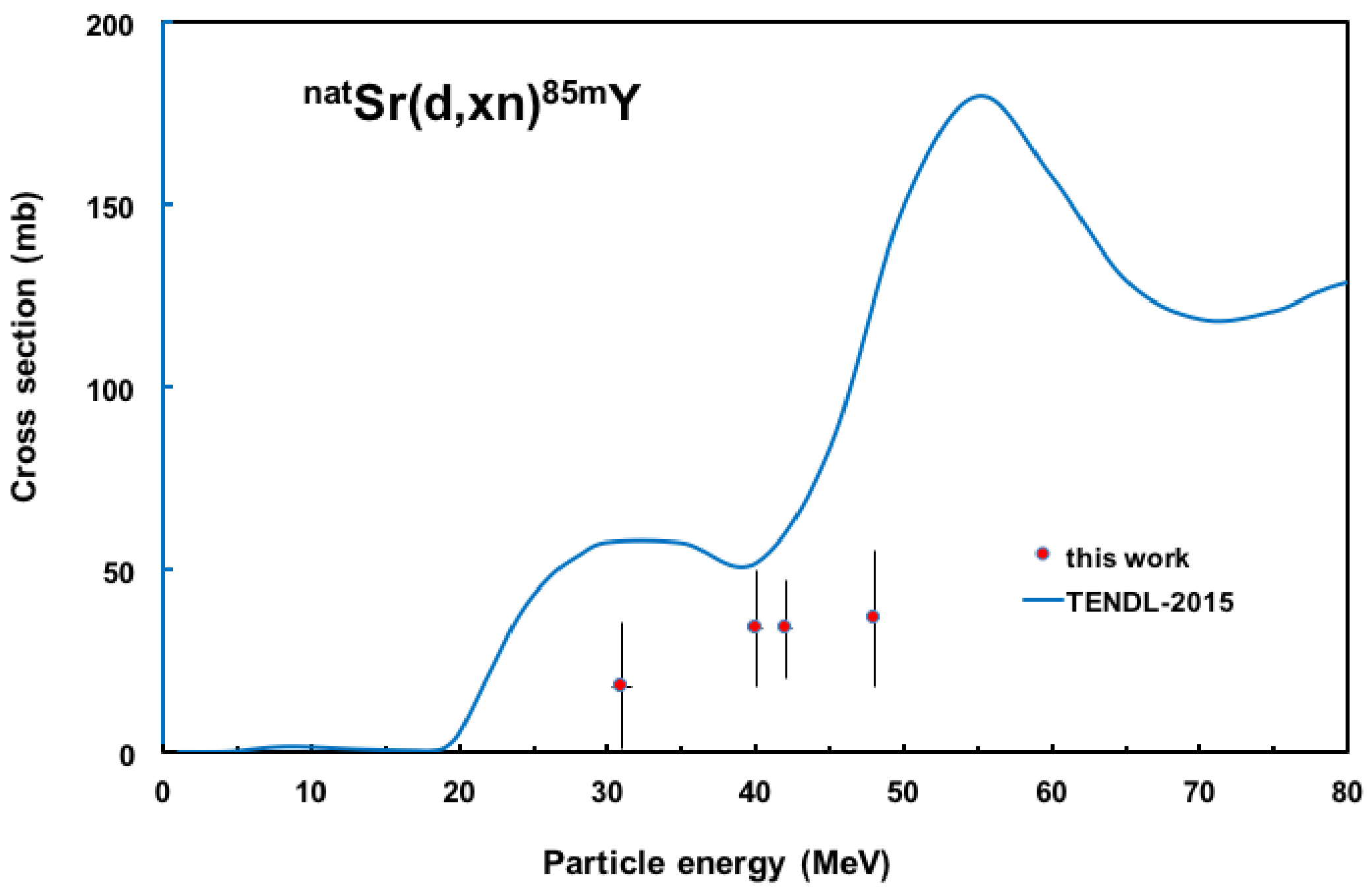

Fig. 5. Excitation function of the $^{nat}Sr(d,x)^{85m}Y$ reaction

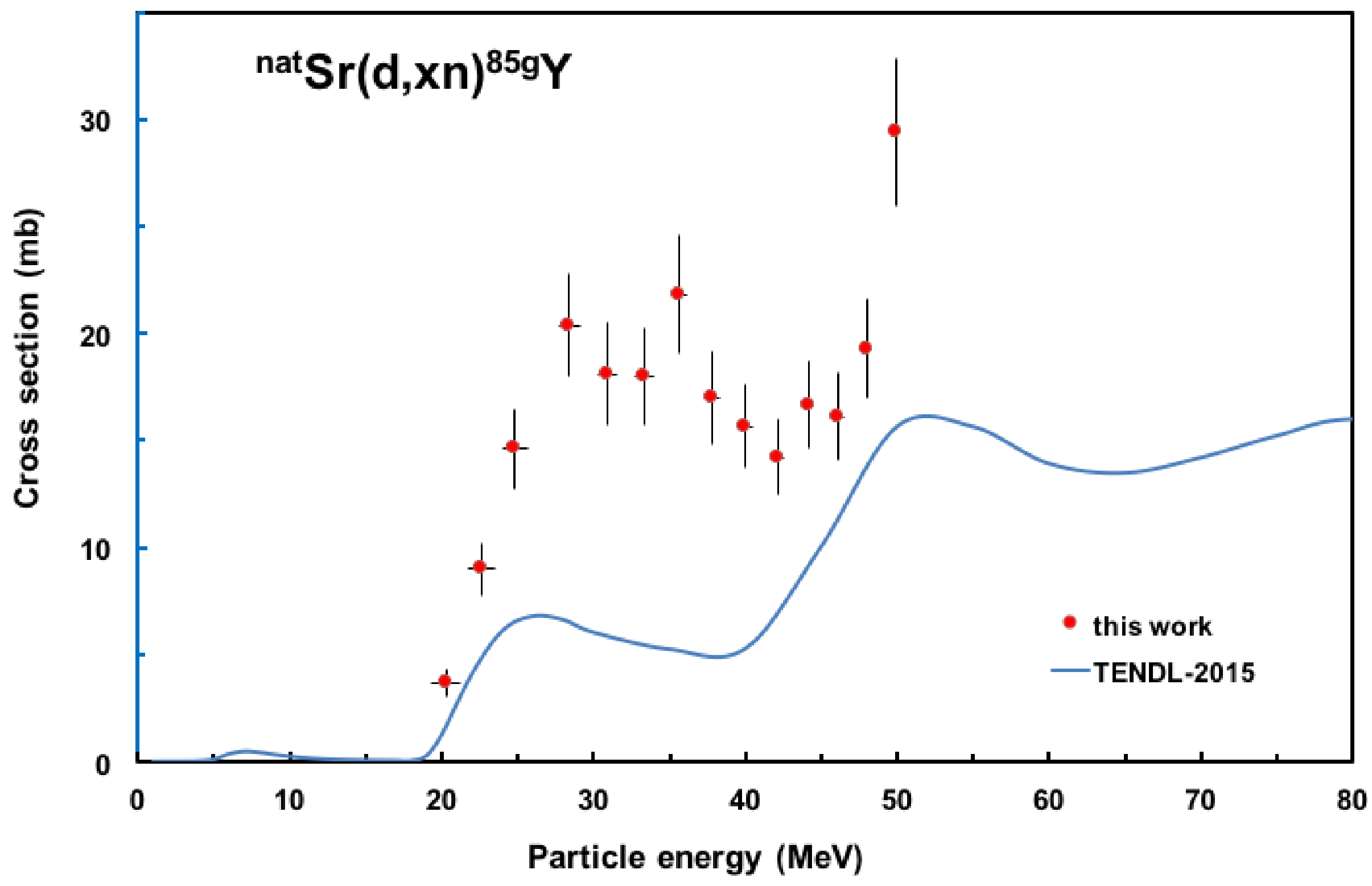

Fig. 6. Excitation function of the $^{nat}Sr(d,x)^{85g}Y$ reaction

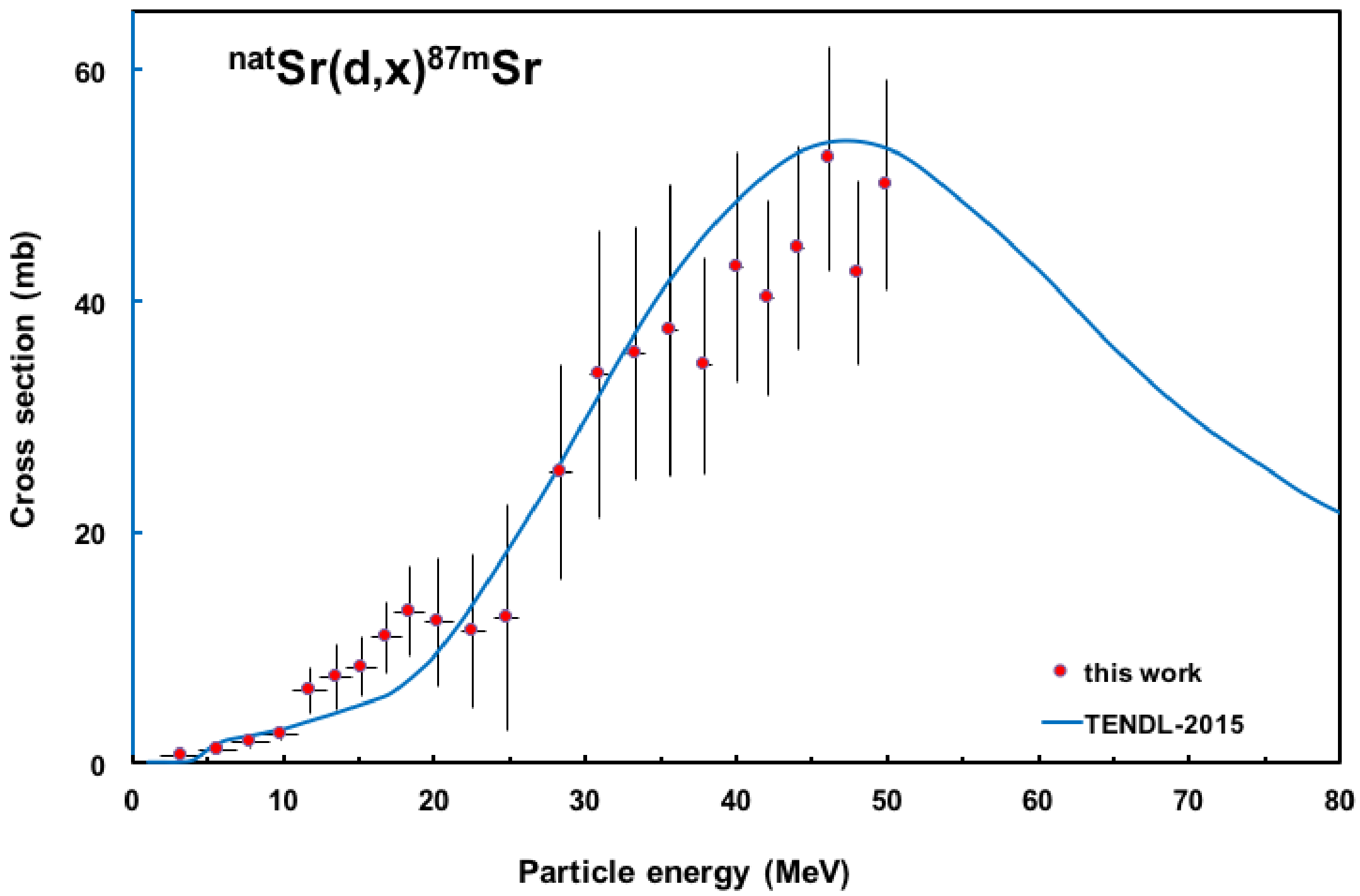

Fig. 7. Excitation function of the $^{nat}Sr(d,x)^{87m}Sr$ reaction

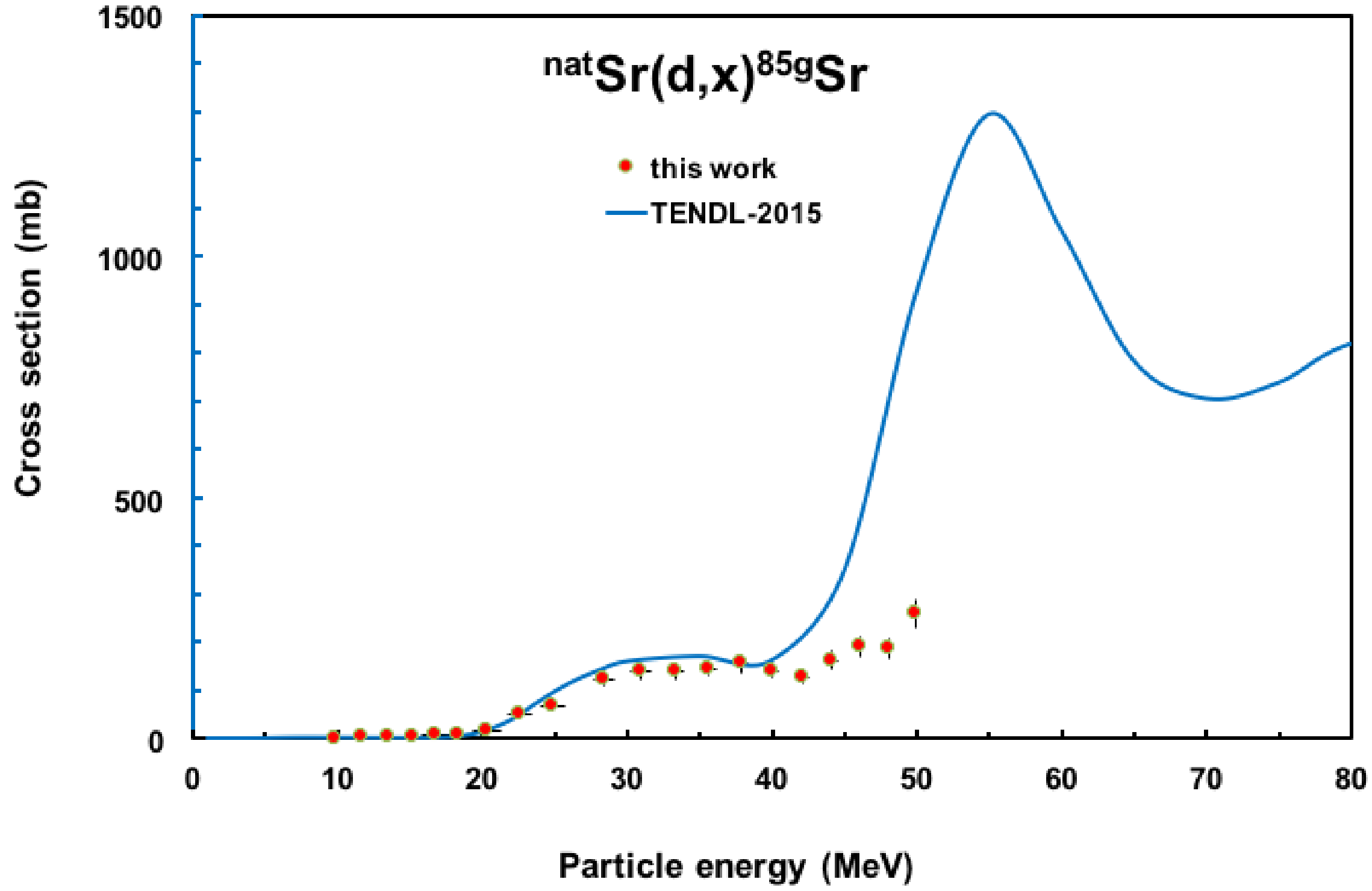

Fig. 8. Excitation function of the $^{nat}Sr(d,x)^{85m}Sr$ reaction

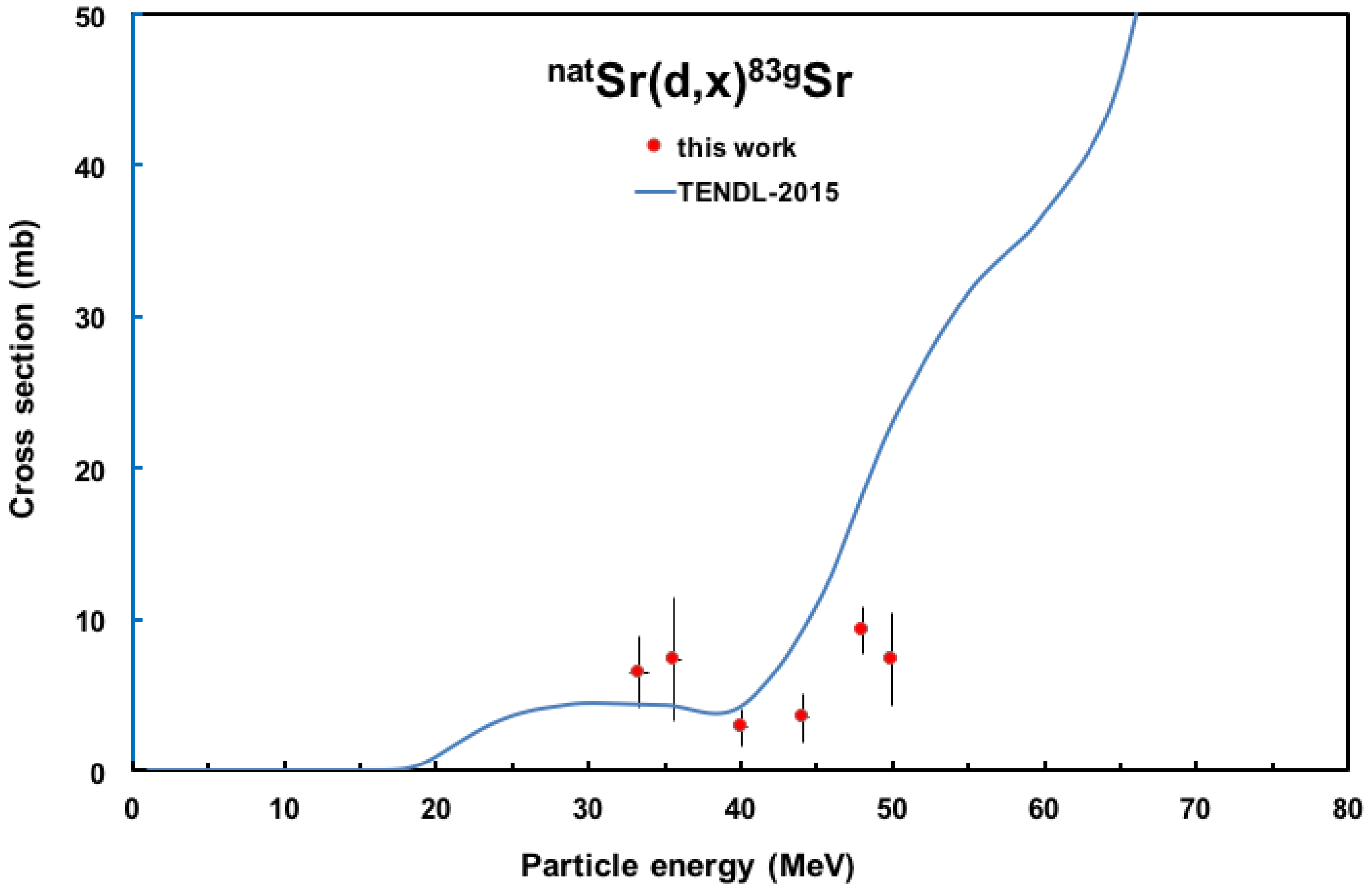

Fig. 9. Excitation function of the $^{nat}Sr(d,x)^{83g}Sr$ reaction

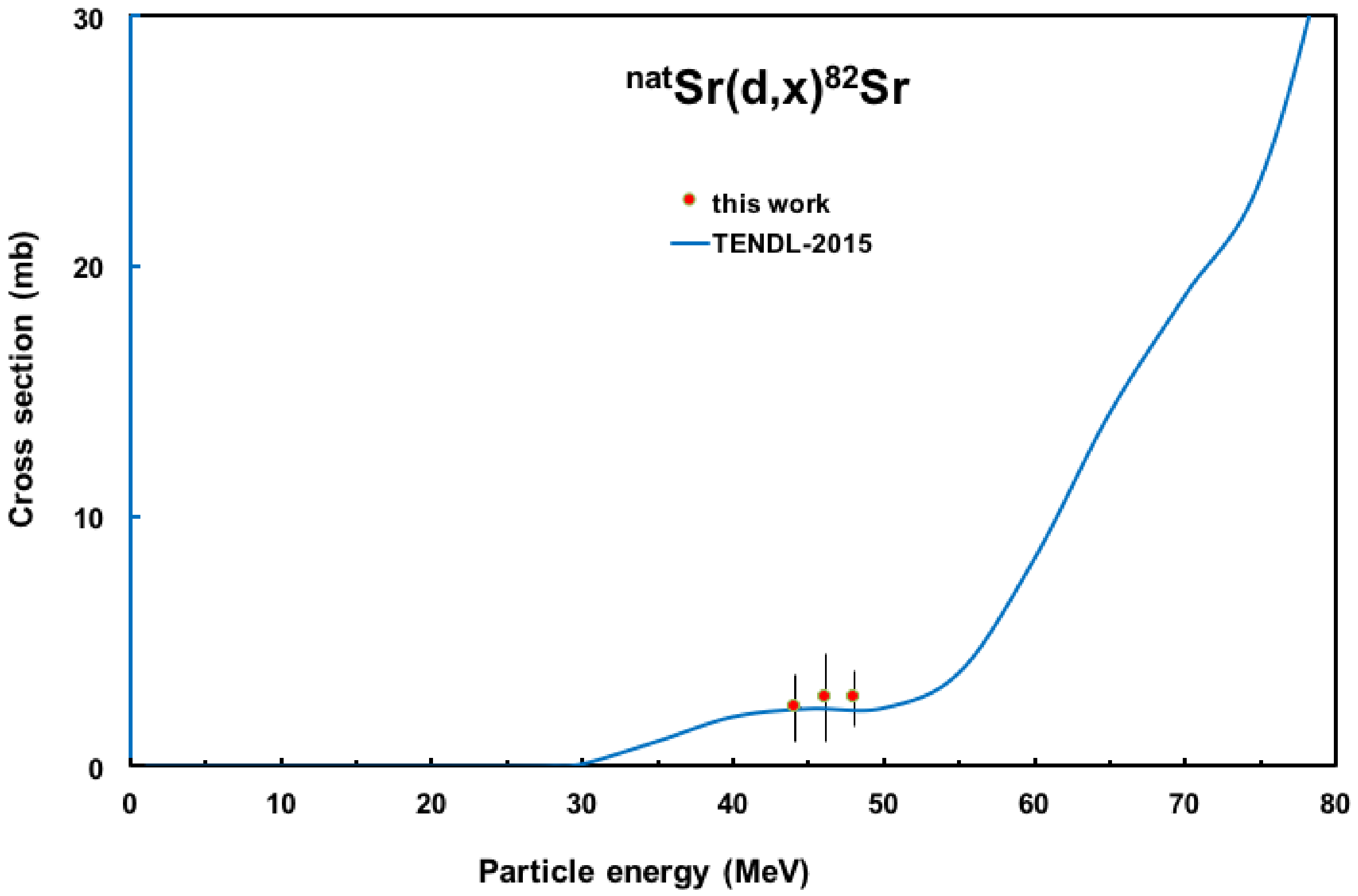

Fig. 10. Excitation function of the $^{nat}Sr(d,x)^{82}Sr$ reaction

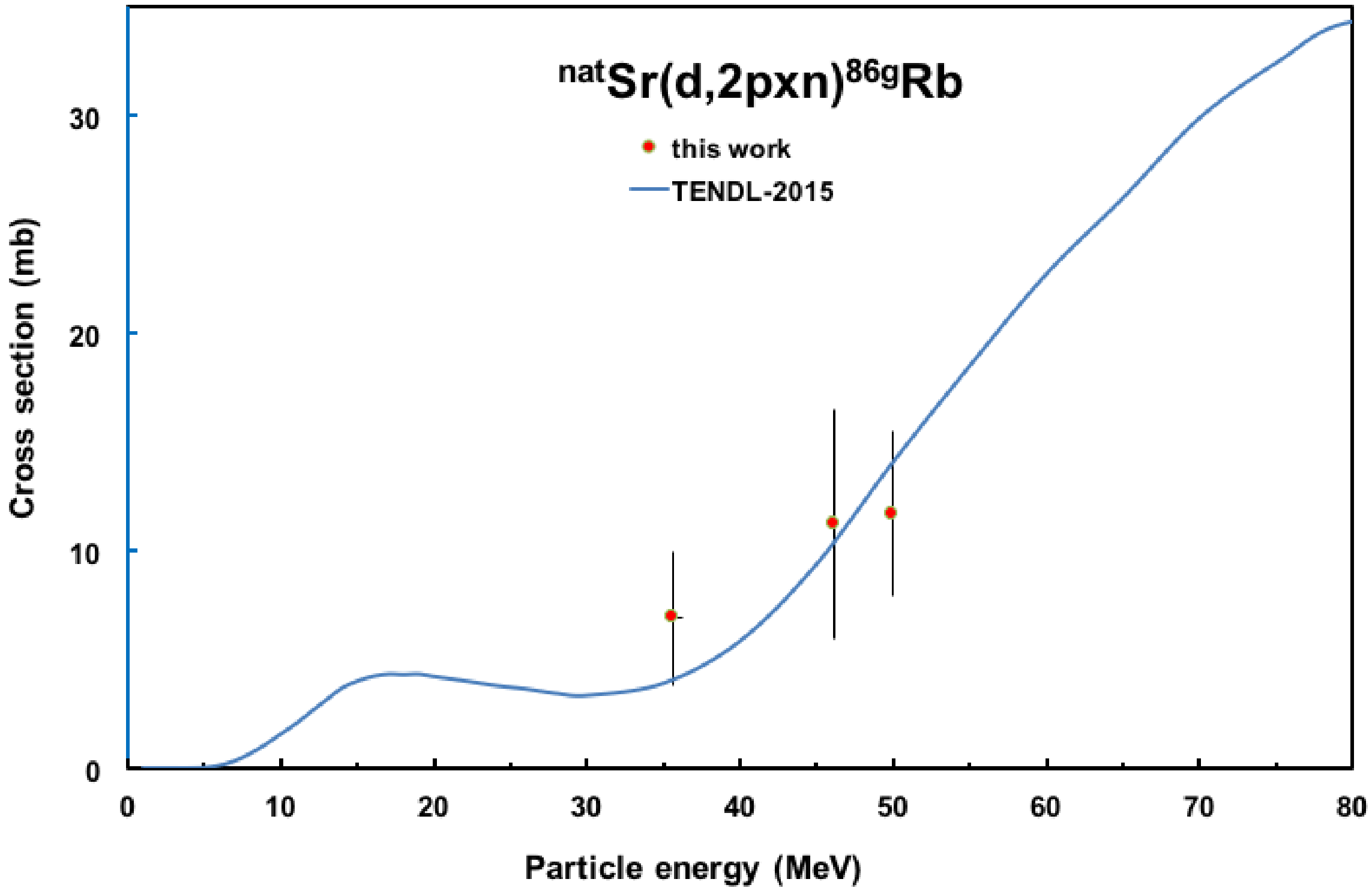

Fig. 11. Excitation function of the $^{nat}Sr(d,x)^{86mg}Rb$ reaction

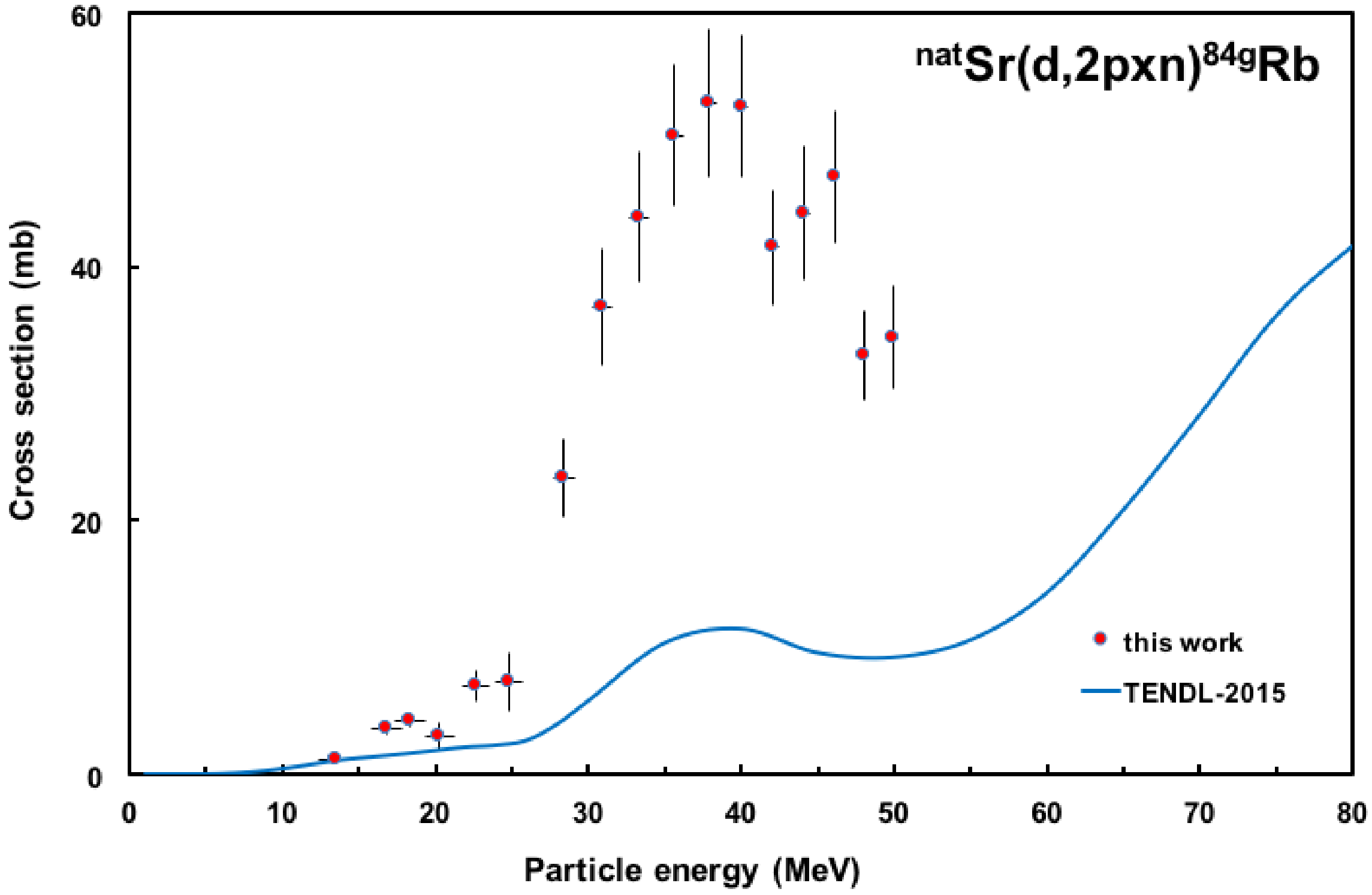

Fig. 12. Excitation function of the $^{nat}Sr(d,x)^{84g}Rb$ reaction

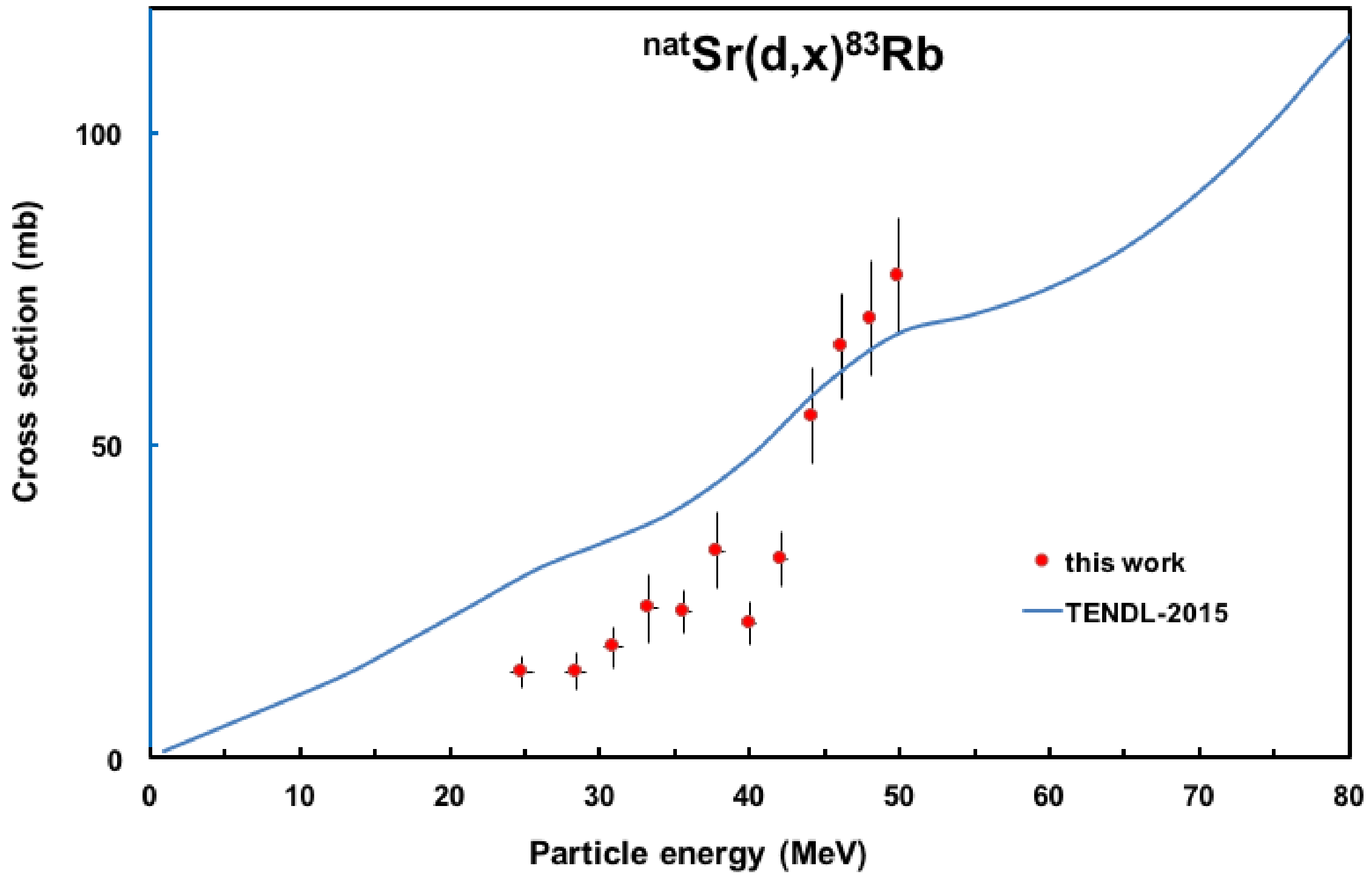

Fig. 13. Excitation function of the ${}^{nat}Sr(d,x){}^{83g}Rb$ reaction

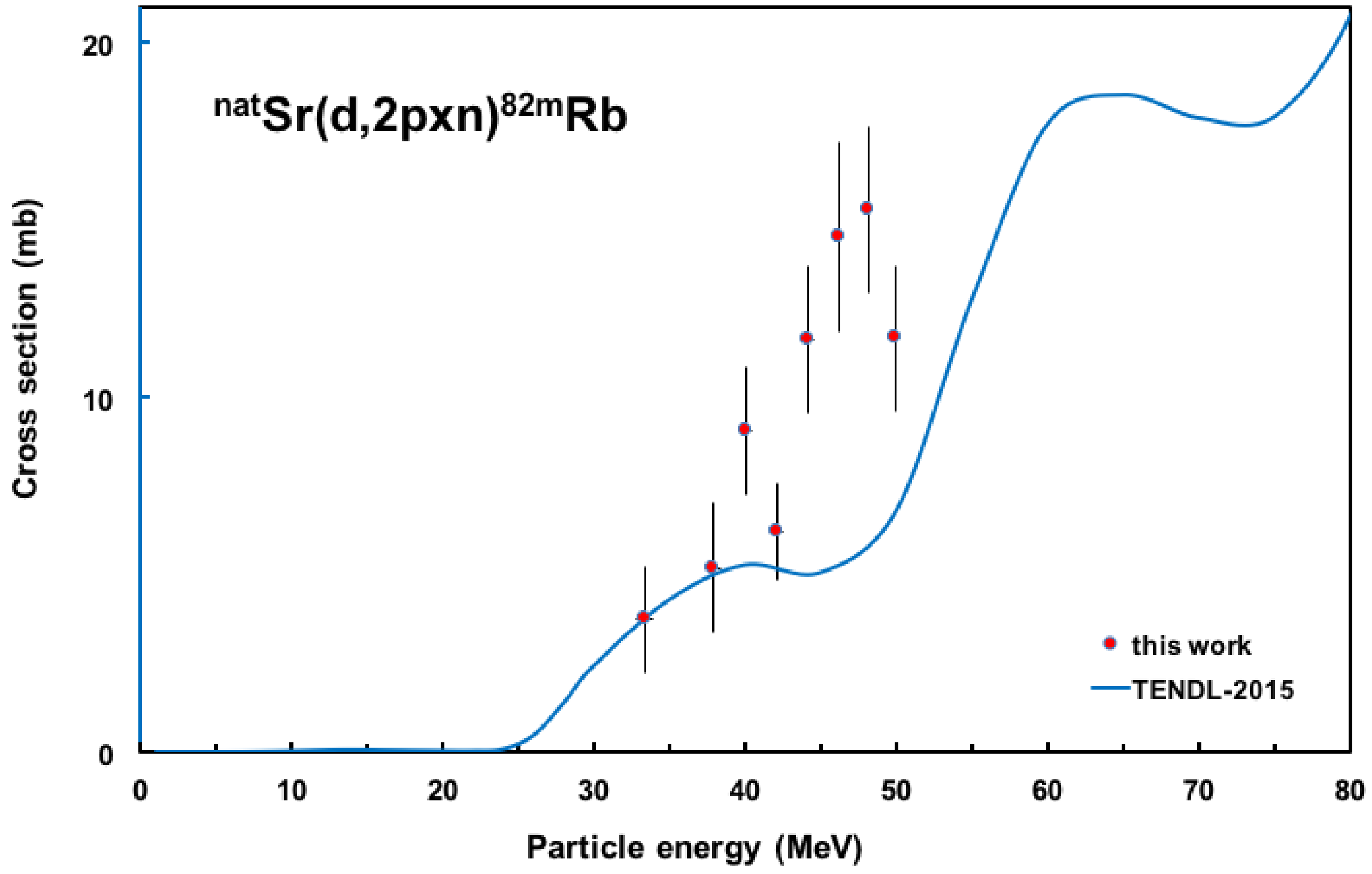

Fig. 14. Excitation function of the ${}^{nat}Sr(d,x){}^{82m}Rb$ reaction

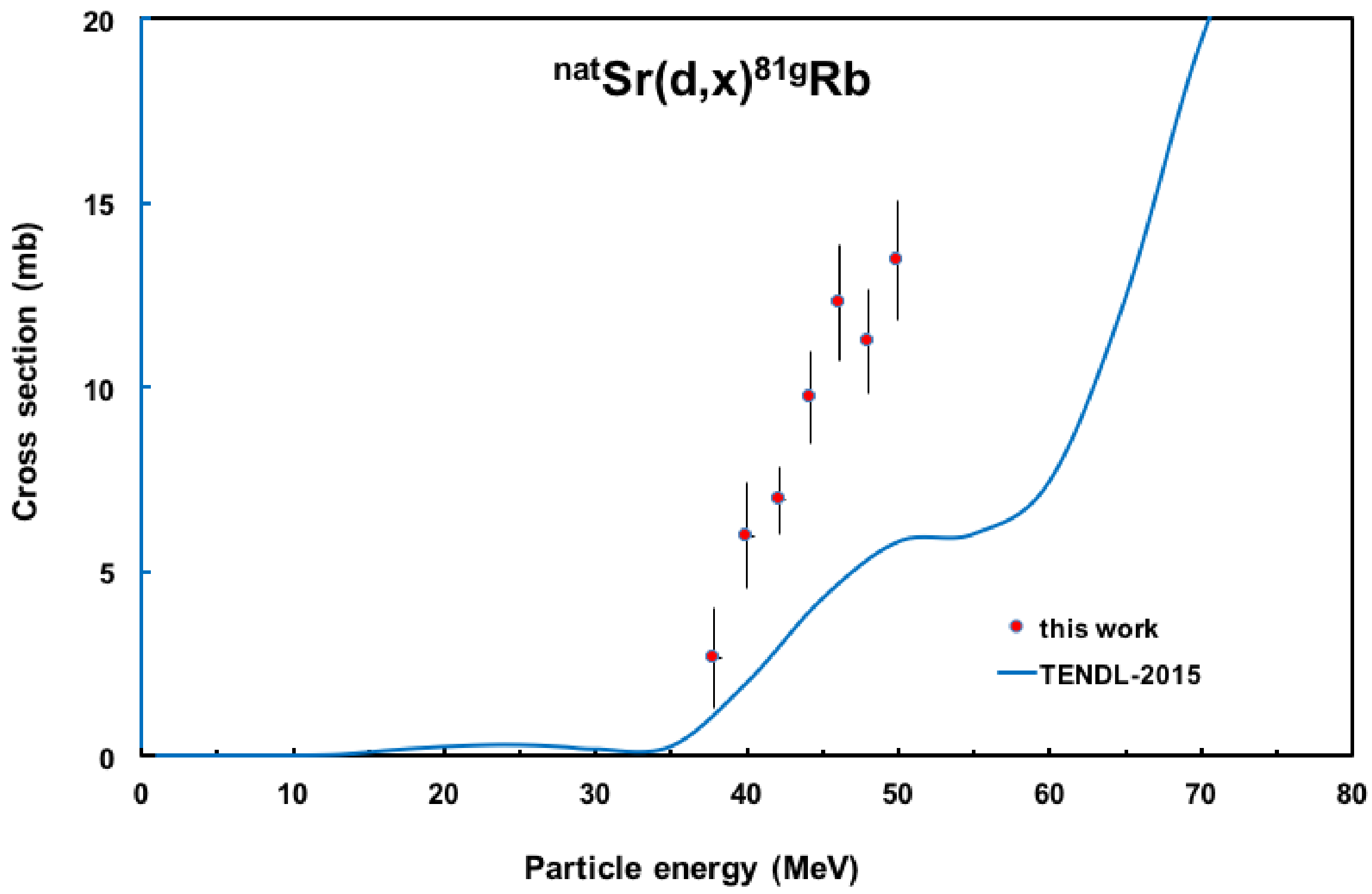

Fig. 15. Excitation function of the $^{nat}Sr(d,x)^{81mg}Rb$ reaction

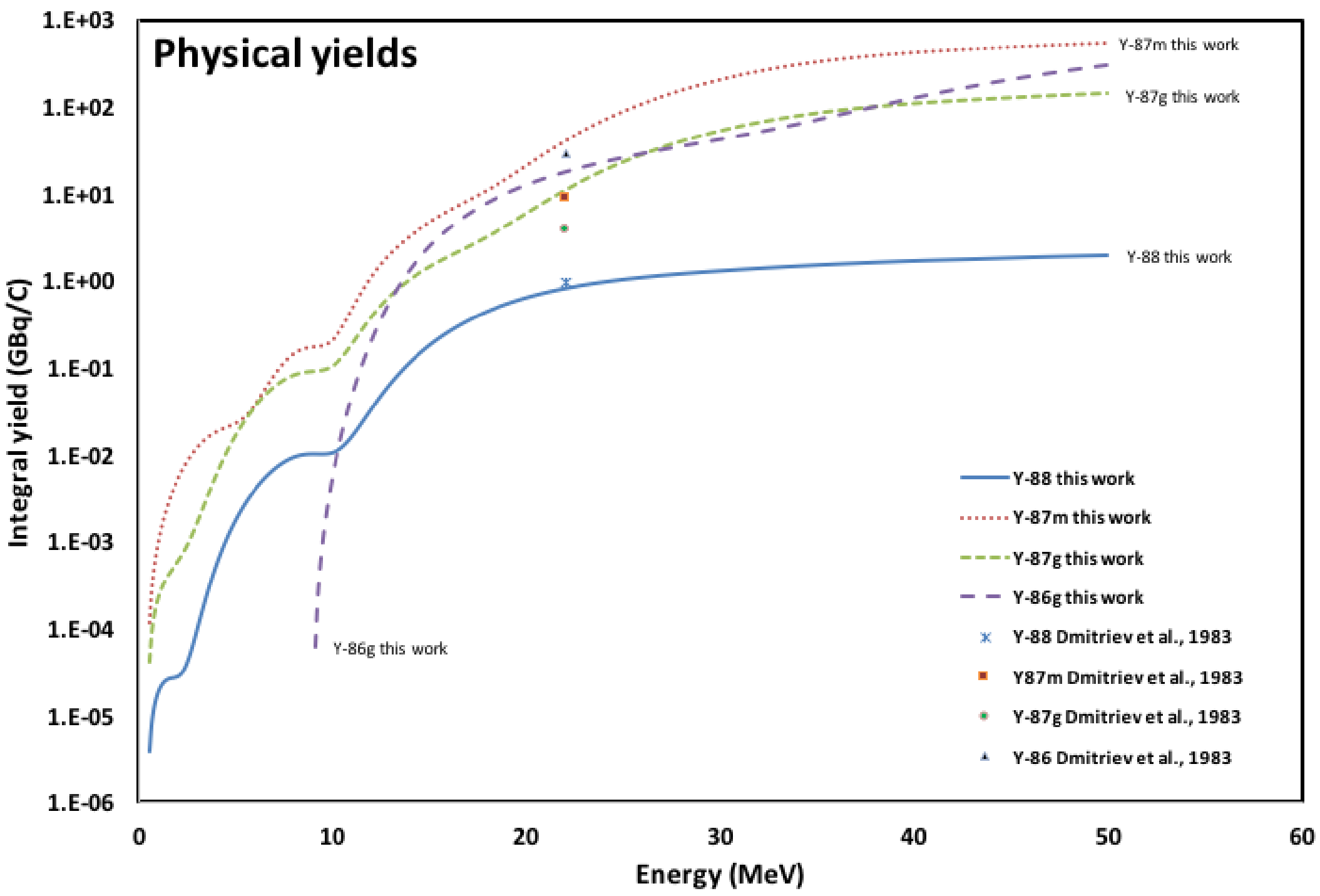

Fig 16. The calculated yields for production of $^{86,87m,87g,88}Y$, compared with the experimental yield

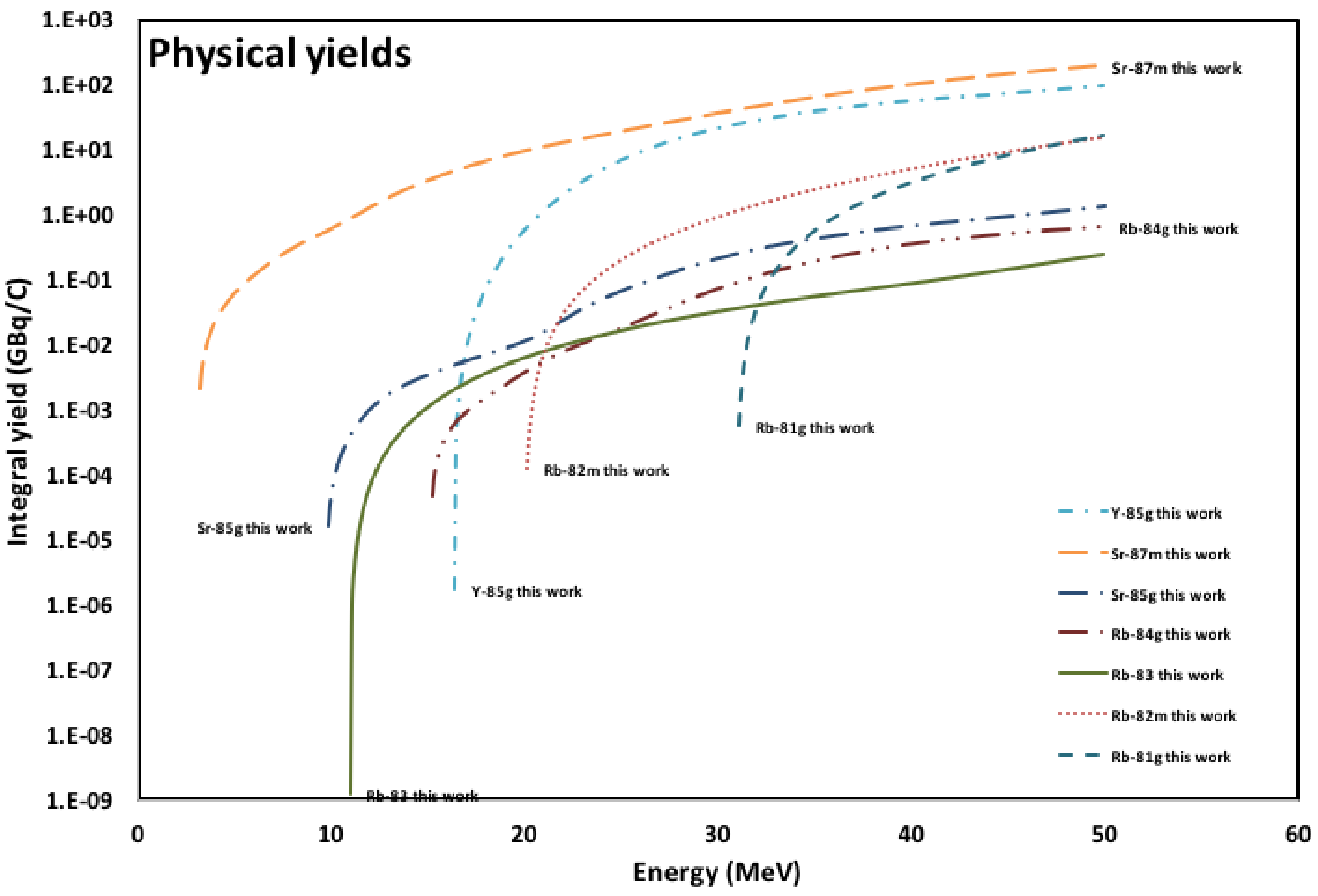

Fig 17. The calculated yields for production of $^{85g}Y$, $^{87m,85g}Sr$ and $^{84g,83,82m,81g}Rb$

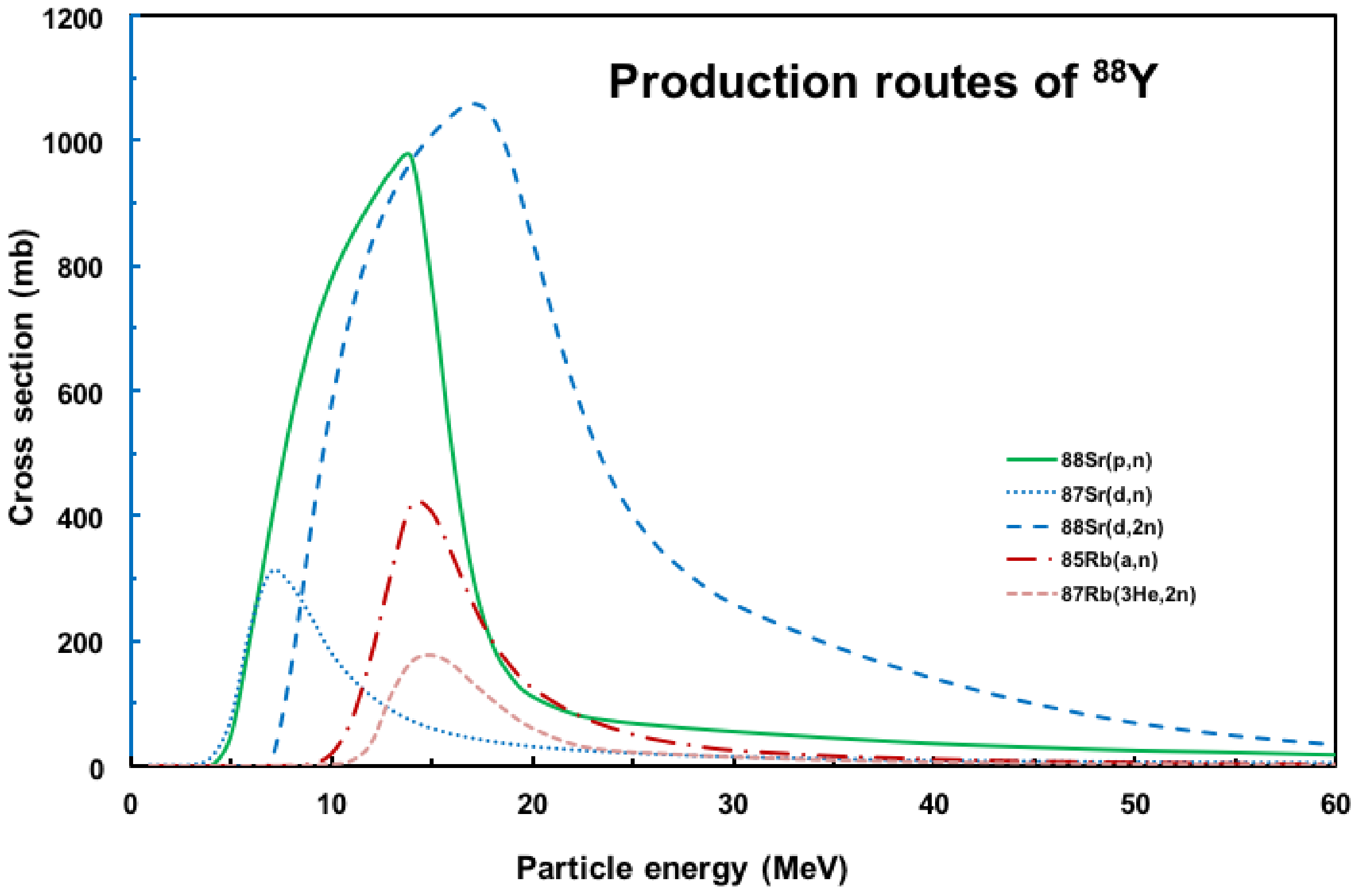

Fig. 18 Comparison of cross-sections (TENDL-2015) for direct production routes of $^{88}Y$

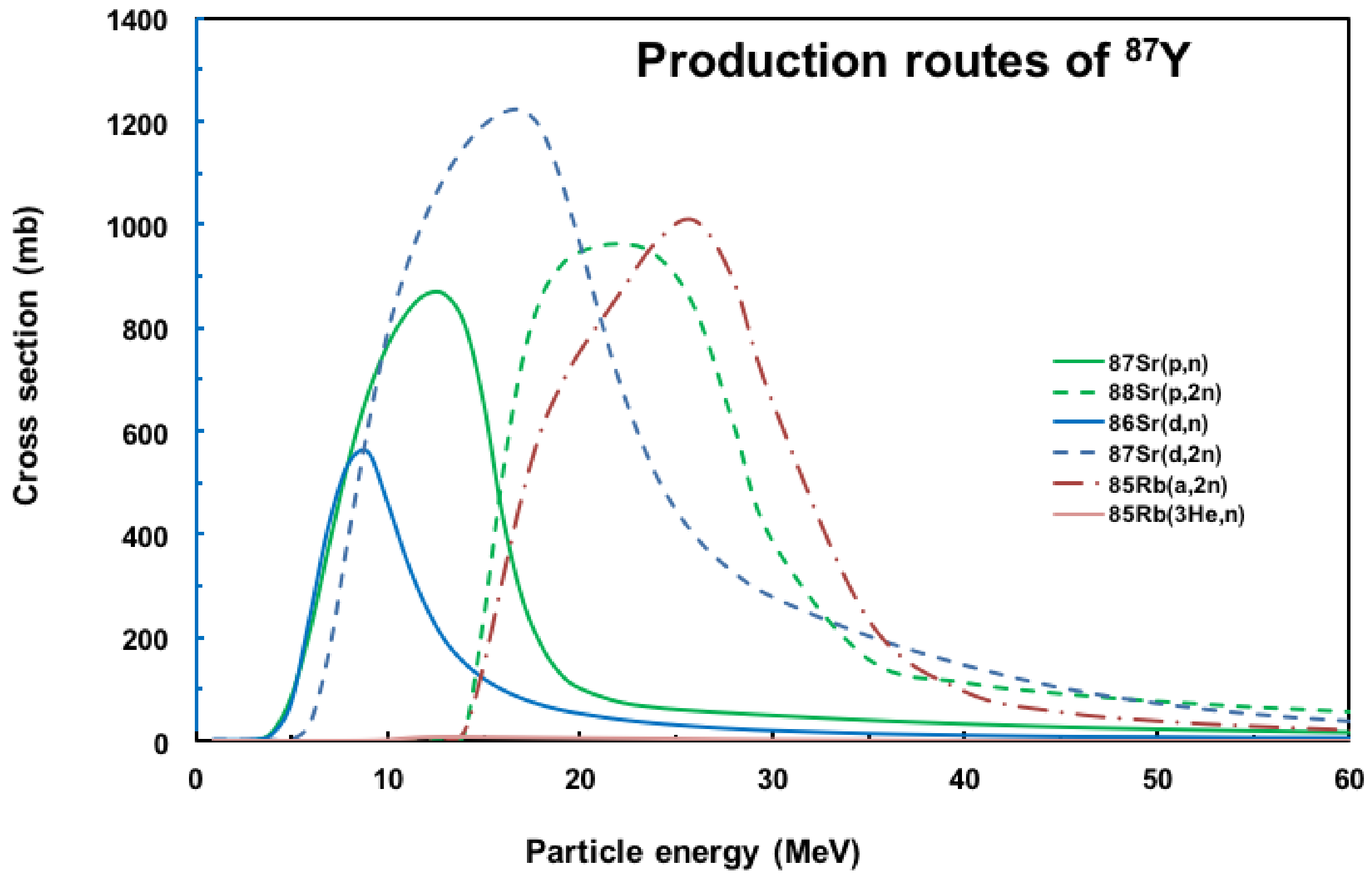

Fig. 19. Comparison of cross-sections (TENDL-2015) for direct production routes of $^{87}$Y

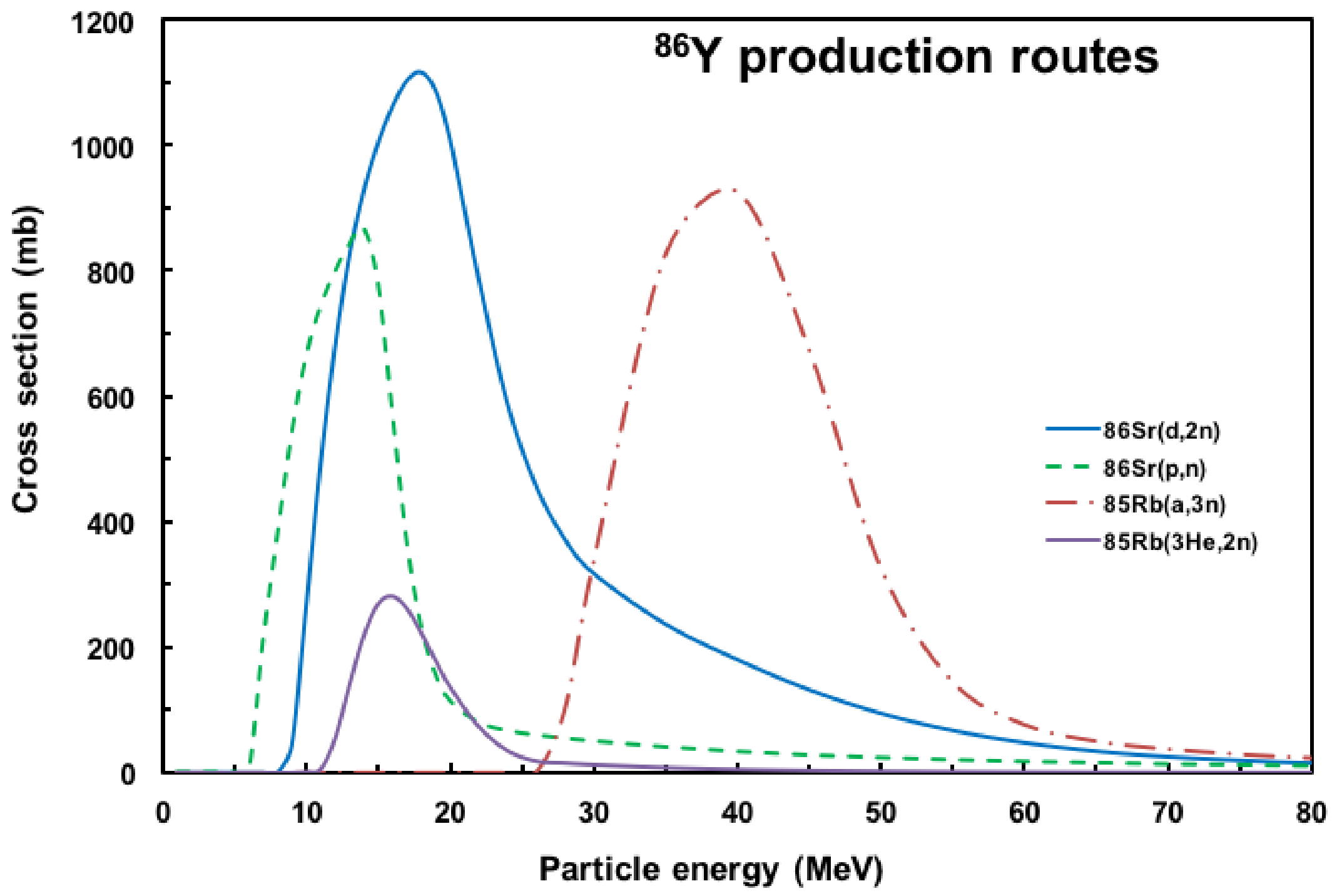

Fig. 20. Comparison of cross-sections (TENDL-2015) for direct production routes of $^{86}$Y

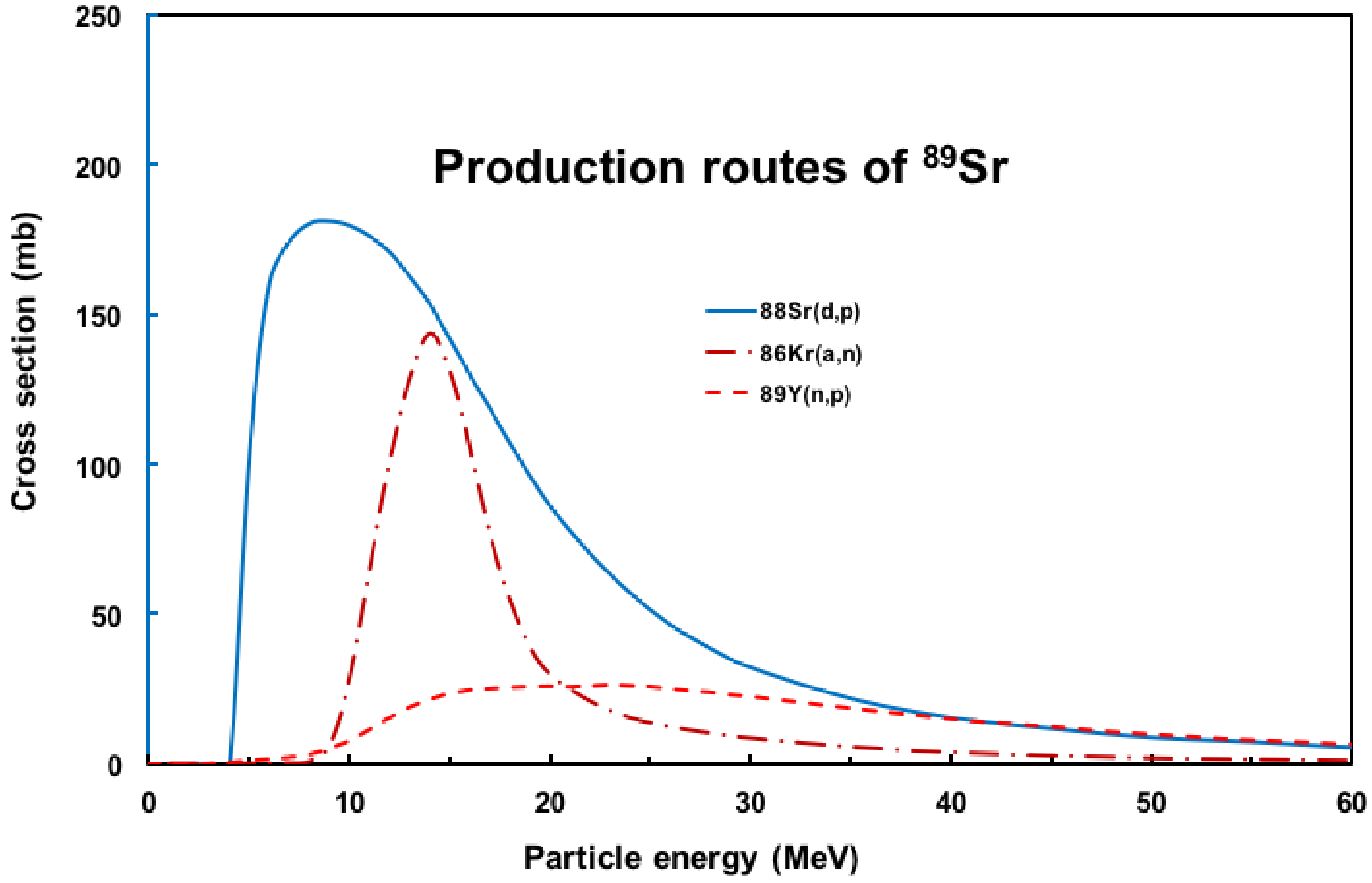

Fig. 21. Comparison of cross-sections (TENDL-2015) for direct production routes of $^{89}Sr$